\documentclass[a4paper,prd,noshowpacs,noshowkeys,notitlepage,nofootinbib]{revtex4}
\pdfoutput=1  
\usepackage[latin1]{inputenc}
\usepackage{amsmath,amssymb,mathrsfs}
\usepackage{graphicx}
\usepackage[]{units}
\usepackage[usenames,dvipsnames]{color} 
\usepackage{hyperref}
\hypersetup{
   colorlinks=true,       
    linkcolor=Blue,          
    citecolor=Plum,        
    filecolor=magenta,      
    urlcolor=YellowOrange           
}
\usepackage[normalem]{ulem} 
\usepackage{setspace}
\providecommand{\lag}{\mathscr{L}}
\providecommand{\tphi}{\tilde{\phi}}
\providecommand{\tla}{\tilde{\lambda}}

\providecommand{\Uz}{U^{(0)}}
\providecommand{\bM}{\bar{M}}

\providecommand{\eps}{\epsilon}
\providecommand{\epsz}{\eps^{(0)}}
\providecommand{\Mp}{M_{\rm pl}}
\providecommand{\cL}{\mathtt{L}}
\providecommand{\cB}{\mathtt{B}}
\providecommand{\om}{\omega}
\providecommand{\cp}{\mathsf{CP}}
\providecommand{\cpmutau}{\cp^{\mu\tau}}
\providecommand{\tm}{\tilde{m}}
\providecommand{\cN}{\mathcal{N}}
\providecommand{\tla}{\tilde{\lambda}}
\providecommand{\yN}[1]{y_{_{N #1}}}
\providecommand{\ZZ}{\mathbb{Z}}
\providecommand{\tp}{{\mss{\mathsf{T}}}}
\providecommand{\mss}[1]{\mbox{\scriptsize $#1$}}
\providecommand{\ml}[1]{\mbox{\large $#1$}}
\providecommand{\eq}[1]{\begin{equation} #1 \end{equation}}
\providecommand{\eqali}[1]{\begin{equation}\begin{aligned} #1
    \end{aligned}\end{equation}}
\providecommand{\subeqali}[2][]{\begin{subequations}\label{#1}\begin{align}
#2    \end{align}\end{subequations}}
\DeclareMathOperator{\re}{\mathrm{Re}} 
\DeclareMathOperator{\im}{\mathrm{Im}} 
\DeclareMathOperator{\diag}{\mathrm{diag}} 
\providecommand{\mtrx}[1]{\begin{pmatrix} #1 \end{pmatrix}}
\providecommand{\ums}[2][1]{\ml{\tfrac{#1}{#2}}} 
\providecommand{\aver}[1]{\langle #1 \rangle}
\providecommand{\mfn}[1]{\mbox{\footnotesize $#1$}}
\usepackage{bbm}
\providecommand{\id}{{\mathbbm{1}}} 
\providecommand{\xlink}[1]
  {\href{http://arxiv.org/abs/#1}{arXiv:#1}}

\begin{document}
\title{
Mu-tau reflection symmetry with a high scale texture-zero
}
\author{C.~C.~Nishi}
\email{celso.nishi@ufabc.edu.br}
\affiliation{
Centro de Matemática, Computação e Cognição\\
Universidade Federal do ABC -- UFABC, 09.210-170,
Santo André, SP, Brazil
}
\author{B. L. S\'anchez-Vega$^{a,b}$}%
\email{bruce.sanchez@ufabc.edu.br}
\author{G.~Souza~Silva$^{a}$}
\email{giovane.s@ufabc.edu.br}
\affiliation{$^{(a)}$%
Centro de Ci\^encias Naturais e Humanas\\
Universidade Federal do ABC -- UFABC, 09.210-170,
Santo Andr\'e-SP, Brasil}
\affiliation{$^{(b)}$%
Instituto de Física Gleb Wataghin -- UNICAMP\\
13083-859 Campinas, São Paulo, Brazil
}

\begin{abstract}
The $\mu\tau$-reflection symmetric neutrino mass matrix can accommodate all known neutrino mixing angles, with maximal atmospheric angle fixed, and predicts all the unknown CP phases of the lepton sector but is unable to predict the absolute neutrino mass scale.
Here we present a highly predictive scenario where $\mu\tau$-reflection is combined with a discrete abelian symmetry to enforce a texture-zero in the mass matrix of the heavy right-handed neutrinos that generate the light neutrino masses.
Such a restriction reduces the free parameters of the low energy theory to zero and 
the absolute neutrino mass scale is restricted to few discrete regions, three in the few meV range and one extending up to around 30 meV.
The heavy neutrino sector is dependent only on two free parameters which are further restricted
to small regions from the requirement of successful leptogenesis.
Mass degenerate heavy neutrinos are possible in one case but there is no resonant enhancement of 
the CP asymmetry.
\end{abstract}
\maketitle
\section{Introduction}

Our picture of the parameters that govern neutrinos physics at low energy are almost complete after the measurement of nonzero reactor angle in 2012\,\cite{dayabay}.
In case neutrinos are Dirac, only the absolute neutrino mass, the mass ordering and one Dirac CP phase is unknown.
The measurement of this CP phase is one of the goals of current experimental efforts to advance our knowledge about neutrinos.
In case neutrinos are Majorana, two more Majorana CP phases should be added to the 
list of unknowns.

One of the simplest symmetries that can predict all the CP phases and yet allow CP 
violation is the symmetry known as $\mu\tau$-reflection symmetry or 
$\cpmutau$ symmetry where the neutrino sector is invariant by exchange of the muon
neutrino with the tau antineutrino\,\cite{mutau-r:HS,mutau-r:GL}; see also review in \cite{mutau-r:review}.
This symmetry predicts maximal Dirac CP phase ($\delta=\pm 90^\circ$) and trivial Majorana phases with discrete choices of the CP parities.
Additionally, the atmospheric angle $\theta_{23}$ is predicted to be maximal
($45^\circ$), well within 2$\sigma$ in the latest global fits\,\cite{global.fit} ($1\sigma$ for normal ordering).
The recent IceCube results on atmospheric neutrinos also corroborate maximal $\theta_{23}$\,\cite{icecube.18}.
Current data also hints at a value of the Dirac CP phase in the broad vicinity of $-90^\circ$.
As a consequence of the symmetry, the fixed values for the CP phases lead to characteristic bands for the possible effective mass of neutrinoless double beta decay but still allows successful leptogenesis\,\cite{yanagida} to occur if flavor effects are taken into account\,\cite{cp.mutau};
see also Ref.\,\cite{lepto.review:ours} for a review on leptogenesis in the presence of flavor symmetries.
If the conditions for maximal atmospheric angle and Dirac CP phase are relaxed, correlations between $\theta_{23}$ and $\delta$ can be tested in the future DUNE and Hyper-K experiments\,\cite{partial.mutau}.
Even the exact $\cpmutau$ case can be tested in DUNE\,\cite{mutau:dune} but 
$\cpmutau$ is too simple to predict the other unknown parameter, i.e, the absolute 
neutrino mass scale.

In that respect, it was shown in Ref.\,\cite{cpmutau:0} that the imposition of an abelian discrete symmetry in conjunction with $\cpmutau$ symmetry could enforce a \textit{one-zero texture} in addition to the $\cpmutau$ form.
Such a setting reduced the number of free parameters in the neutrino mass matrix from five to four to account for the four observables $\Delta m^2_{21},\Delta m^2_{32},\theta_{12},\theta_{13}$ -- the rest are fixed from symmetry -- and a highly predictive scenario emerged where the absolute neutrino mass was fixed and further correlations of parameters appeared. 
Texture-zeros in the lepton sector were first studied in Ref.\,\cite{frampton} and can be systematically obtained with the imposition of abelian symmetries\,\cite{texture0:abelian}.
In the original proposal of the $\mu\tau$ interchange symmetry\,\cite{fukuyama}, a similar approach of enforcing one texture-zero was also adopted to increase predictivity;
see also Refs.\,\cite{mutau-inter}.
Generically two texture-zeros are still allowed by data\,\cite{2-zero} but our combined approach only allows for one because $\cpmutau$ relates some entries.
In this approach, the abelian symmetry cannot be arbitrary as well because it should 
satisfy certain consistency conditions\,\cite{holthausen} to be combined with 
$\cpmutau$.
It was shown in Ref.\,\cite{cp.mutau} that the smallest $\ZZ_n$ that can be combined nontrivially\,\footnote{%
Excluding the most common $g\to g$ and $g\to g^{-1}$ automorphisms.
}
with $\cpmutau$ is $\ZZ_8$.
Note that this setting of $\cpmutau$ and $\ZZ_8$ is much simpler than embedding $\cpmutau$
and an abelian symmetry in larger nonabelian discrete symmetries\,\cite{mutau:r:models,real.sym}.\footnote{%
Sometimes, one of them can be accidental.
In any case, when not accidental, the flavor group should be enlarged to include $\mathsf{CP}$.
}
The use of nonabelian discrete symmetries to describe the lepton flavor structure has been extensively analyzed\,\cite{GF:review}.

Here, we propose a modified but equally predictive setting where the texture zero appears instead at the high scale, in the mass matrix of the right-handed neutrino in the context of the simple type I seesaw.
This texture zero will be directly transmitted to the \textit{inverse} of the light neutrino mass matrix due to the seesaw form when the neutrino Dirac mass matrix is diagonal\,\cite{lavoura}.\footnote{%
A similar idea was explored in Ref.\,\cite{lavoura:mutau} in the context of $\mu\tau$ interchange symmetry.}
Therefore, the light neutrino mass matrix will still depend on four parameters and the
predictive power of the low energy theory is the same as in Ref.\,\cite{cpmutau:0}.
However, since this setting comes directly from a high scale model, the structure of the heavy neutrinos will be also highly constrained.
One of the key byproducts of the seesaw mechanism ---the possibility to generate the baryon asymmetry of the Universe through leptogenesis\,\cite{yanagida}--- can be studied 
and the few free parameters can be constrained from the requirement of successful 
leptogenesis.
This differs from other ways of increasing predictivity in the context of leptogenesis such as imposing texture zeros in the different mass matrices in the minimal case of two right-handed neutrinos\,\cite{lepto:min:texture} or considering larger flavor symmetries\,\cite{lepto:GF}.

The outline of the paper is as follows: in Sec.\,\ref{sec:model} we present the model and show how the texture-zero at high scale is generated.
Section \ref{sec:nu} analyzes the predictions for the parameter space for light neutrinos arising from the model.
The heavy neutrino spectrum and mixing pattern that only depend on two parameters are analyzed in Sec.\,\ref{sec:heavy}.
Constraints from successful leptogenesis on the parameter space are discussed in Sec.\,\ref{sec:lepto}.
Finally, the conclusions are presented in Sec.\,\ref{sec:concl} and the appendices contain some auxiliary material.

\section{$\cpmutau$ with high scale texture-zero}
\label{sec:model}

In the context of the type I seesaw mechanism where the light neutrino mass matrix $M_\nu$ is related to the inverse of the heavy neutrino mass matrix $M_R$ by $M_\nu=-M_D^\tp M_R^{-1}M_D$, it is not difficult to envisage that texture-zeros in $M_R$ can lead to texture-zeros in the \textit{inverse} of the light neutrino mass matrix when the neutrino Dirac mass matrix $M_D$ is diagonal\,\cite{lavoura}.
Applying this idea, we will show here that it is possible to have a $\cpmutau$ symmetric neutrino mass matrix together with a texture-zero in the \textit{inverse} matrix $M_\nu^{-1}$.
The latter will have the form
\eq{
\label{Mnui:cpmutau}
M_\nu^{-1}=
\left(
\begin{array}{ccc}
 a & d & d^* \\
 d & c & b \\
 d^* & b & c^* \\
\end{array}
\right)\,,\quad
\text{with real $a,b$ and $\im(d^2 c^*)\neq 0$}\,,
}
with phenomenologically viable texture-zeros in the $(ee)$ or $(\mu\tau)$ entries, i.e., $a=0$ or $b=0$, respectively.
The high predictivity of this setting will be analogous to Ref.\,\cite{cpmutau:0} 
and the absolute neutrino mass scale can be fixed to discrete values.
Additionally, since this setting comes from a high scale model, the structure of the heavy neutrinos will be also highly constrained and leptogenesis can be studied.
Only two free parameters will control the heavy neutrino sector.

The defining property of a complex symmetric matrix $A$ which is $\cpmutau$ symmetric is
\eq{
\label{cpmutau:A}
X^\tp A X=A^*\,,
\quad
\text{with}
\quad
X=\mtrx{1&0&0\cr0&0&1\cr0&1&0}\,.
}
Clearly $A=M_\nu^{-1}$ in \eqref{Mnui:cpmutau} satisfies this property and once satisfied, it is also valid for its inverse $A^{-1}=M_\nu$, i.e., the neutrino mass matrix will also have the form \eqref{Mnui:cpmutau}.
A $\cpmutau$ symmetric Majorana neutrino mass matrix can be enforced by $\cpmutau$ at the level of fields acting as\,\cite{mutau-r:GL}
\eq{
\label{cp:nu}
\nu_{e L}\to \nu_{eL}^{cp}
\,,\quad
\nu_{\mu L}\to \nu_{\tau L}^{cp}
\,,\quad
\nu_{\tau L}\to \nu_{\mu L}^{cp}
\,,
}
where $cp$ denotes the usual CP conjugation.\footnote{%
Note that a relative global phase is not relevant and e.g. a mass matrix antisymmetric by $\mu\tau$-reflection\,\cite{samanta} leads effectively to the same consequences.
}
Additionally, we will use the same $\ZZ_8$ of the previous paper \cite{cpmutau:0} acting on charged leptons as
\eq{
\label{def:T}
e\sim -1
\,,\quad
\mu\sim \om_8
\,,\quad
\tau\sim \om_8^3
\,,\quad
~~\om_8=e^{i2\pi/8}.
}
It was shown in Ref.\,\cite{cp.mutau} that $\ZZ_8$ was the minimal abelian symmetry where a nontrivial combination with $\cpmutau$ is possible.

We can think that these two symmetries ---$\ZZ_8$ and $\cpmutau$--- initially act on the left-handed lepton doublets $(L_e,L_\mu,L_\tau)$ before 
they are spontaneously broken.
Then the two symmetries act on the same space and $\cpmutau$ induces the 
following automorphism on $\ZZ_8$\,\cite{cp.mutau}:
\eq{
T\to XT^*X^{-1}=T^5\,,
}
where $T$ encodes the $\ZZ_8$ transformation in \eqref{def:T} and $X$ denotes $\nu_\mu$-$\nu_\tau$ interchange in \eqref{cpmutau:A}.
We also note that the rephasing transformations that preserve $\ZZ_8$ in 
\eqref{def:T} and $\cpmutau$ in \eqref{cp:nu} are of the form
\eq{
\label{rephasing}
L_e\to \pm L_e,\quad
L_\mu\to e^{i\alpha}L_\mu,\quad
L_\tau\to e^{-i\alpha}L_\tau.
}
It is clear that these transformations also preserve the form of the mass matrix in 
\eqref{Mnui:cpmutau} and can be used to make $c$ or $d$ real.
Flavor independent rephasing by $i$ also preserves the form of the mass matrix 
(flips the sign of $a,b$) but changes $\cpmutau$ by a global sign.
Hence, only the relative sign of $a$ and $b$ is significant.

In the charged lepton sector, 
the $\mu\tau$ mass difference arises from a large source of $\cpmutau$ breaking at high energy\,\cite{cp.mutau}; see appendix \ref{ap:mutau}
for more details.
After that stage, the $\ZZ_8$ will remain as a residual symmetry so that we are simply left with
\eq{
\label{lag:l}
-\lag^l= h_e\bar{L}_e\phi l_{eR}+h_\mu\bar{L}_\mu\phi l_{\mu R}+h_\tau\bar{L}_\tau\phi l_{\tau R}
\,.
}
We assume that the physics responsible for such a $\cpmutau$ breaking is well above the scale
of the heavy neutrinos which come from $\ZZ_8$ breaking.

Light neutrino masses will come from the type I seesaw mechanism where we add three 
singlet neutrinos $N_{\alpha R}$, $\alpha=e,\mu,\tau$.
The $N_{\alpha R}$ and left-handed lepton doublets $L_\alpha$ transform under 
$\ZZ_8$ and $\cpmutau$ in the same way as in eqs.\,\eqref{def:T} and \eqref{cp:nu}.
So the neutrino Dirac mass matrix will be diagonal.

To avoid bare terms, we also introduce a $\ZZ_4^{B-L}$ symmetry under which the 
lepton doublets $L_\alpha$ and the singlet neutrinos $N_{\alpha R}$ carry charge 
$-i$. Heavy neutrino masses will be generated by singlet scalars $\eta_k$ with 
$\ZZ_4^{B-L}$ charge $-1$.
Each of $\eta_k$ carries a charge $\om_8^k$ of $\ZZ_8$ and then $\eta_0,\eta_4$ can be real.
The fields $\eta_1$ and $\eta_3$ are necessarily present and are connected by 
$\cpmutau$ as
\eq{
\eta_1\to \eta_3^*\,.
}
The rest of the fields, $\eta_2,\eta_0,\eta_4$, transform trivially under $\cpmutau$\,\cite{cpmutau:0}.

Then the neutrino Yukawa couplings will be
\eqali{
-\lag_N &= \yN{e}\bar{N}_{eR}\tphi L_e
+\yN{\mu}\bar{N}_{\mu R}\tphi L_\mu +\yN{\tau}\bar{N}_{\tau R}\tphi L_\tau
\cr&~~
+\ \ums{2}c_{ee}\eta_0 \bar{N}_{eR}N^c_{eR}
+ \ums{2}c_{\mu\mu}\eta_2 \bar{N}_{\mu R}N^c_{\mu R}
+ \ums{2}c_{\tau\tau}\eta_2^* \bar{N}_{\tau R}N^c_{\tau R}
\cr&~~
+\ c_{e\mu}\eta_3^* \bar{N}_{e R}N^c_{\mu R}
+ c_{e\tau}\eta_1^* \bar{N}_{e R}N^c_{\tau R}
+ c_{\mu\tau}\eta_4 \bar{N}_{\mu R}N^c_{\tau R}
+h.c.,
}
where, due to $\cpmutau$, $\yN{e},c_{ee}$ and $c_{\mu\tau}$ are real while
$\yN{\tau}=\yN{\mu}^*$, $c_{\tau\tau}=c_{\mu\mu}^*$ and $c_{e\tau}=c_{e\mu}^*$.

The Dirac mass matrix will be diagonal as
\eq{
M_D=v\diag(\yN{e},\yN{\mu},\yN{\tau})
=m_D\diag(1,\kappa,\kappa^*)
\,,
}
where $m_D=v\,\yN{e}$ is real by symmetry and $\kappa=|\kappa|$ can be made real and 
positive by rephasing $L_\alpha$.
The heavy neutrino mass matrix will have the $\cpmutau$ symmetric form
\eq{
\label{MR:param}
M_R=\mtrx{A & D & D^* \cr \star & C & B \cr \star & \star & C^*}\,,
}
where e.g. $A=c_{ee}\aver{\eta_0}$.
We assume that $\cpmutau$ is preserved by $\eta_k$, i.e., 
\eq{
\label{eta:cpc}
\aver{\eta_1}=\aver{\eta_3}^*\,.
}

Light neutrino masses will be generated by the seesaw mechanism as $M_\nu= -M_D^\tp M_R^{-1} M_D$, whose inverse is closely related to $M_R$ as
\eqali{
\label{Mnu:-1}
M_\nu^{-1}&=-M_D^{-1}M_R M_D^{\tp -1}
\cr
&=-m_D^{-2}
\mtrx{A & \kappa^{-1} D & (\kappa^{-1} D)^* \cr
	\star & \kappa^{-2}C & |\kappa|^{-2}B \cr
	\star & \star & (\kappa^{-2} C)^*}
=\mtrx{a & d & d^* \cr \star & c & b \cr \star & \star & c^*}
\,.
}
We get the texture-zero $a=0$ or $b=0$ if either $\eta_0$ or $\eta_4$ is absent and that is inherited from texture-zeros in $M_R$ in the same positions ($A=0$ or $B=0$).
When solutions exist to accommodate the oscillation data, the matrix $M_\nu^{-1}$ is completely fixed, except for experimental error.
We show the possible solutions in Sec.\,\ref{sec:nu}.
And then, $M_R$ will depend only on two free parameters, $m_D,\kappa$, as
\eq{
\label{MR}
M_R=-M_DM_\nu^{-1}M_D^\tp=
-m_D^2
\mtrx{a & \kappa d & (\kappa d)^* \cr \star & \kappa^2c & |\kappa|^2b \cr \star & \star & (\kappa^2c)^*}
\,.
}
We will use $m_D$ or $\yN{e}$ interchangeably as one of the free parameters.

Concerning mixing angles, it is guaranteed that any matrix in the form \eqref{Mnui:cpmutau}, which is symmetric by $\cpmutau$, can be always
diagonalized by a matrix of the form\,\cite{mutau-r:HS,mutau-r:GL}
\eq{
\label{U0}
\Uz=\mtrx{u_1&u_2&u_3\cr
	w_1& w_2 & w_3\cr
	w_1^*& w_2^* & w_3^*}\,,
}
where $u_i$ are all real and positive.
Moreover, the Majorana type diagonalization (also known as Takagi factorization) will already lead to a real diagonal matrix and only discrete choices of signs ---the CP parities--- will appear instead of Majorana phases.
In this way, the mass matrices for the light and heavy neutrinos can be diagonalized as
\subeqali[Mnu.MR:U]{
\label{Mnu:U}
{\Uz_\nu}^\tp M_\nu \Uz_\nu&=\diag(m_i')
\,,
\\
\label{MR:U}
{\Uz_R}^\dag M_R{\Uz_R}^*&=\diag(M_i')
\,,
}
where $\Uz_\nu$ and $\Uz_R$ are in the form \eqref{U0},
and the primed masses denote $m_i'=\pm m_i$ and $M_i'=\pm M_i$, with $m_i$ and $M_i$ being 
the actual light and heavy masses.
The complex conjugation in $\Uz_R$ appears because it is defined as the transformation 
matrix for $N_R$ whereas $M_R$ is defined in the basis $N_R^cN_R^c$.
So Eq.\,\eqref{Mnu.MR:U} implies that the full diagonalizing matrices can be written as
\eq{
\label{U=U0.K}
U_\nu=\Uz_\nu K_\nu\,,
\quad
U_R=\Uz_R K_R\,,
}
where $K_\nu,K_R$ are diagonal matrices of $1$ or $i$ depending on the signs on 
\eqref{Mnu:U} or \eqref{MR:U}, respectively.
Since a sign flip of \textit{both} $M_\nu$ and $M_R$ is not physical, we can distinguish four discrete cases of CP parities according to the sign of the diagonal entries of $K_\nu^2$\,\cite{cp.mutau} as
\eq{
\label{cp.parities}
K_\nu^2:\quad
(+++),~(-++),~(+-+),~(++-)\,.
}
As we seek texture-zeros, some cancellation between $m'_i$ will be necessary and the case $(+++)$ will not appear in our solutions.
The generic possibilities for $K_R^2$ as well as the detailed mass spectrum and mixing pattern will be discussed in Sec.\,\ref{sec:heavy}.
Opposite parities in $K^2_R$ will also give rise to cancellations in the CP asymmetries 
of heavy neutrinos suppressing the resonant enhancement.

We limit ourselves here to discussing briefly the limit $\kappa=1$, which is straightforward.
Considering \eqref{MR} and since 
\eq{
\label{diag:inv}
{\Uz_\nu}^\dag M_\nu^{-1} {\Uz_\nu}^*=\diag(m_i'^{-1})
\,,
}
we can identify
\eq{
\label{UR:k=1}
\Uz_R={\Uz_\nu}\,.
}
With this equation fixing the ordering for $(M_1',M_2',M_3')$ in \eqref{MR:U}, we have the direct relation
\eq{
\label{Mi:mi:k=1}
M_i'=-\frac{m_D^2}{m_i'}\,.
}
This means that the spectrum for the heavy neutrinos is completely fixed in terms of the light masses and the CP parities for the heavy neutrinos are opposite to those of the light neutrinos. Therefore, $K_R^2=-K_\nu^2$ and
\eq{
U_R=iU_\nu\,.
}

As $\kappa$ deviates from unity, $\Uz_R$ will deviate from ${\Uz_\nu}$
 depending only on the parameter $\kappa$. The same will happen for the mass ratios between two heavy masses.
Only the absolute scale for $M_i$ will be controlled by $m_D$ (or $\yN{e}$).

\section{Light neutrinos}
\label{sec:nu}

The inverse of the light neutrino mass matrix in the flavor basis is $\cpmutau$ symmetric and was given in \eqref{Mnui:cpmutau} with $a$ or $b$ possibly vanishing.
Different texture-zeros are not phenomenologically possible because it would lead to vanishing $\theta_{13}$ (or also $\theta_{12}$)\,\cite{cpmutau:0}.
Since $M_\nu$ itself is $\cpmutau$ symmetric, the usual predictions of maximal $\theta_{23}=45^\circ$ and $\delta=\pm 90^\circ$ follow as $\theta_{13}\neq 0$\,\cite{mutau-r:HS,mutau-r:GL}.

Without texture-zeros, the five parameters in \eqref{Mnui:cpmutau} ---$a,b,|c|,|d|,\arg(d^2c^*)$---
should describe the remaining five observables not fixed by symmetry: 
$\theta_{12},\theta_{13},m_1,m_2,m_3$.
Among these five observables, only four combinations are currently experimentally 
determined and we cannot predict the only unknown quantity: the lightest neutrino mass 
(equivalently, the absolute neutrino mass scale). With the additional 
one-zero texture, the number of free parameters is reduced by one and all the observables 
can be fixed, including the lightest neutrino mass.
We show the possible solutions in table~\ref{table.results} when we allow for the 
experimental uncertainties for observables not fixed by symmetry, in accordance to the 
global-fit in Ref.\,\cite{capozzi.17}\,\footnote{%
More up to date fits are available \cite{global.fit} but the variation is small within 3$\sigma$ ranges.
}.
The procedure to find these solutions are explained below.
A relatively wide range for $m_1$ appears for case II because it is a merger of two 
discrete solutions that would appear if there were no experimental error.
\begin{table}[h]
$\begin{array}{|c|c|c|c|c|c|c|}
\hline
\text{Case} & \mfn{(M_\nu^{-1})_{\alpha\beta}\!=\!0} & \text{ordering} & \text{CP 
parities} & m_0 
 & m_{\beta\beta} & \sum m_\nu \\
\hline
\text{I} & (\mu\tau) & {\rm NO} & (-++) & 2.48\text{ -- }4.36 & 1.25\text{ -- }1.93 
 & 60.7\text{ -- }66.3 \\
\text{II} & (\mu\tau) & {\rm NO} & (+-+) & 4.28\text{ -- }27.31 & 1.84\text{ -- 
}14.42 & 
    63.3\text{ -- }114.6 \\
\text{III} & (ee) & {\rm IO} & (-++) & 1.86\text{ -- }4.27 & 
     13.48\text{ -- }24.77 & 99.7\text{ -- }107.1 \\
\text{IV} & (ee) & {\rm IO} & (++-) & 0.943\text{ -- }1.27 & 
     47.49\text{ -- }50.14 & 98.7\text{ -- }103.7 \\        
\text{V} & (\mu\tau) & {\rm IO} & (++-) & 154\text{ -- }183 & 
    154\text{ -- }182 & 476\text{ -- } 563
\\ \hline
\end{array}
$
\caption{\label{table.results}
Possibilities for one-zero textures with predictions for the lightest 
neutrino mass ($m_0$), neutrinoless double beta decay effective mass 
($m_{\beta\beta}$) and sum of neutrino masses; all masses are in \unit{meV}.
}
\end{table}

\begin{figure}[h]
\includegraphics[scale=0.45]{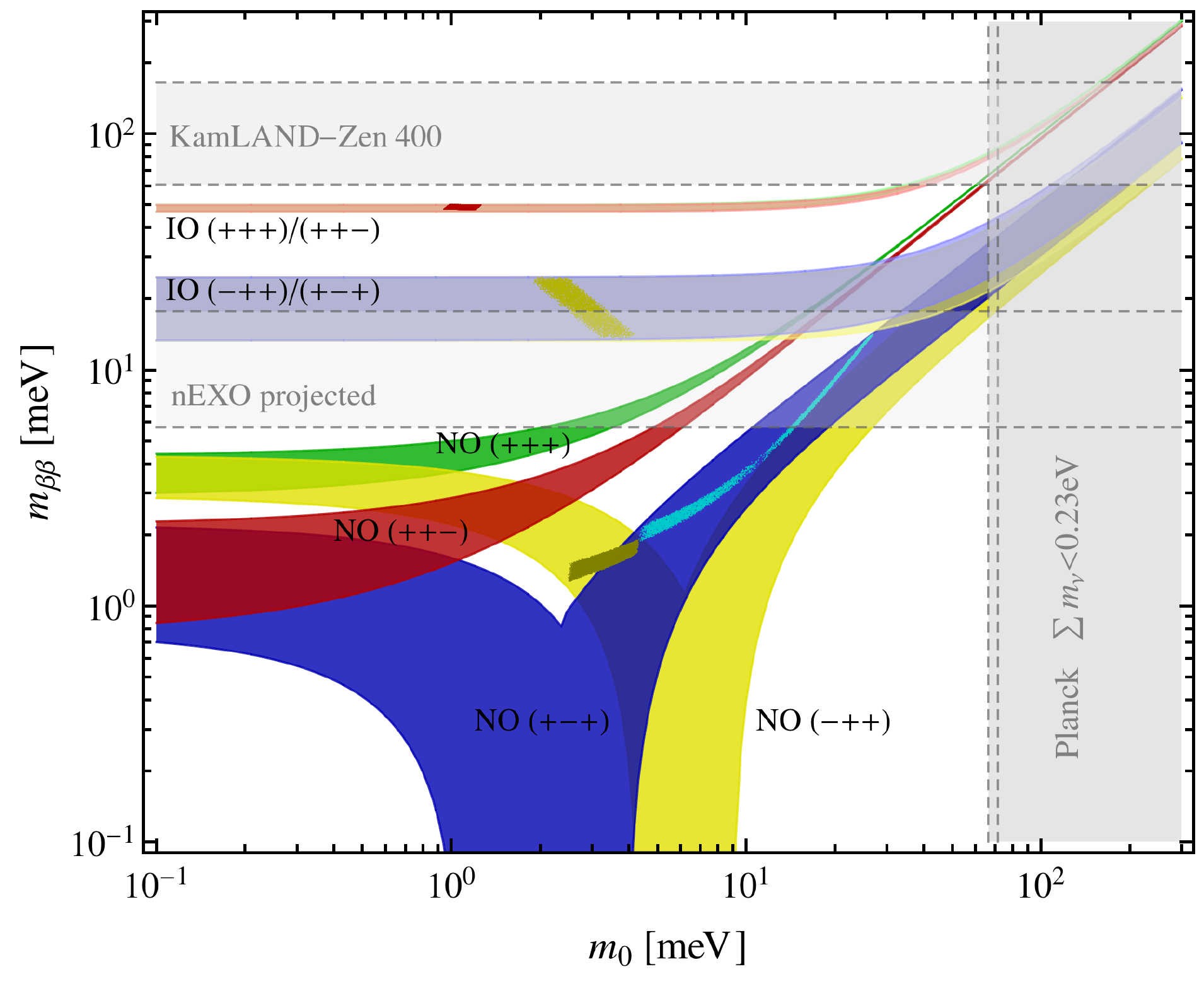}
\caption{\label{fignu.mee}
The colored strips indicate
$m_{\beta\beta}$ as a function of the lightest neutrino mass $m_0$ for generic 
$\cpmutau$ neutrino mass matrix for different cases of CP parities and mass 
orderings\,\cite{cp.mutau}.
Small darker (lighter for blue) regions inside the colored strips mark the 
solutions for our cases IV, III, II and  I (from top to bottom) shown in table 
\ref{table.results}.
}
\end{figure}

We can see that case V has too large masses and it is excluded by the Planck power spectrum limit (95\% C.L.)\,\cite{planck},
\eq{
\sum_{i}m_i< \unit[230]{meV}.
}
We are left with two cases for the normal ordering (NO) and two cases for the inverted 
ordering (IO).
All these cases are also compatible with the latest KamLAND-Zen upper limit for the 
neutrinoless double beta decay parameter at 90\%C.L.\,\cite{kamland-zen},
\eq{
\label{kamland-zen}
m_{\beta\beta}<(61\text{ -- }165)\,\unit{meV}\,.
}
The variation in the latter, comes from the uncertainty in the various evaluations 
of the nuclear matrix elements.
In the near future, experiments such as KamLAND-Zen 800 will probe the IO region that 
includes our case IV.
To see the discovery potential, we show in Fig.\,\ref{fignu.mee} the solutions for cases I, II, III and IV with possible values of $m_{\beta\beta}$ as a function of the lightest mass $m_0$
overlapped with the strips of the generic case with $\cpmutau$ but without any texture-zero\,\cite{cp.mutau}.
We also show the current bounds from KamLAND-Zen 400 in \eqref{kamland-zen} and the future projected sensitivity of the nEXO experiment at 90\% C.L.\,\cite{nexo}.
If this experiment reaches such a sensitivity, it will certainly probe our case III completely and our case II partially.

The solutions in table \ref{table.results} are obtained with the expressions for $a,b$ in terms of physical parameters, which we show below.
To derive them, we first choose the parametrization for the PMNS matrix, without Majorana phases, as
\eq{
\label{U0:param}
\Uz_\nu=\left(
\begin{array}{ccc}
 1 & 0 & 0 \\
 0 & \frac{1}{\sqrt{2}} & \frac{- i}{\sqrt{2}} \\
 0 & \frac{1}{\sqrt{2}} & \frac{+i}{\sqrt{2}} \\
\end{array}
\right)
\mtrx{ c_{13}&0&s_{13}\cr
    0&1&0\cr
    -s_{13}&0&c_{13}}
\mtrx{ c_{12}&s_{12}&0\cr
    -s_{12}&c_{12}&0\cr
    0&0&1}
\,,
}
where, e.g., $c_{13}=\cos\theta_{13}$, and we are choosing the Dirac CP phase to be $e^{i\delta}=-i$ following the current hints from global fits\,\cite{capozzi.17}; the opposite Dirac CP phase can be used by taking the complex conjugate of \eqref{U0:param}.
Note that the standard parametrization corresponds to 
$\diag(1,1,-1)\Uz_\nu\diag(1,1,+i)$.
The parametrization in \eqref{U0:param} obeys the $\cpmutau$ symmetric form \eqref{U0}
but with the additional rephasing freedom from the left fixed by the choice $\re(\Uz_\nu)_{\mu 3}=0$ and $\re(\Uz_\nu)_{\mu 2}>0$\,\cite{cpmutau:0}.
This phase convention implies a certain phase relation between $c$ and $d$ in \eqref{Mnui:cpmutau}.
With that phase convention in mind,
\eqref{Mnu:U} is still guaranteed\,\cite{mutau-r:GL}.

If we invert the relation \eqref{diag:inv} by using \eqref{U0:param}, we can write the parameters $a,b,c,d$ in terms of the neutrino inverse masses and 
mixing angles: 
\eqali{ 
\label{params}
a&=c^2_{13}(m_1'^{-1}c^2_{12} +m_2'^{-1}s^2_{12})+m_3'^{-1} s^2_{13}\,, 
\cr 
b&=\ums{2}\big[m_1'^{-1}s^2_{12}+m_2'^{-1}c^2_{12} +s^2_{13}(m_1'^{-1}c^2_{12}+m_2'^{-1}s^2_{12}) 
+m_3'^{-1} c^2_{13}\big]\,, 
\cr 
d&=\frac{c_{12}s_{12}c_{13}}{\sqrt{2}}(m_2'^{-1}-m_1'^{-1}) 
+i\frac{s_{13}c_{13}}{\sqrt{2}}\big[-m_3'^{-1}+m_1'^{-1}c^2_{12}+m_2'^{-1}s^2_{12}\big]\,, 
\cr 
c&=\ums{2}\big[m_1'^{-1}(s^2_{12}-c^2_{12}s^2_{13})+m_2'^{-1}(c^2_{12}-s^2_{12}s^2_{13})
-m_3'^{-1}c^2_{13}\big] 
+i\,c_{12}s_{12}s_{13}(m_2'^{-1}-m_1'^{-1})\,. 
} 
Choosing $e^{i\delta}=+i$ instead, would correspond to taking $d\to d^*$ and $c\to c^*$.
Note that the phases of $c,d$ in \eqref{params} follow a specific phase relation characterized by the compatibility between
\eq{
\tan\theta_{13}=\frac{\im c}{\sqrt{2}\re d}>0
\quad
\text{and}
\quad
\tan2\theta_{13}=\frac{2\sqrt{2}\im d}{a-b+\re c}\,,
}
necessary for the consistency of \eqref{diag:inv}.
The rephasing freedom in \eqref{rephasing} changes the phases of $c$ and $d$ accordingly.
Other relations between the parameters in \eqref{Mnui:cpmutau} and the physical parameters 
can be extracted from Ref.\,\cite{cpmutau:0} by replacing $m_i'\to m_i'^{-1}$ and 
$\Uz_\nu\to {\Uz_\nu}^*$.
For example, a rephasing invariant measure of CP violation is given by 
\eq{
\label{c*d2}
\im(c^*d^2)=\ums{2}s_{13}c^2_{13}s_{12}c_{12}
\Big(\frac{1}{m_1'}-\frac{1}{m_2'}\Big)
\Big(\frac{1}{m_2'}-\frac{1}{m_3'}\Big)
\Big(\frac{1}{m_3'}-\frac{1}{m_1'}\Big)
\,,
}
which is nonzero in all physical cases.
We would obtain the same result with opposite sign if we had $e^{i\delta}=i$.

Finally, with the expressions for $a$ and $b$ in hand, we can seek solutions for 
$a=0$ or $b=0$ depending on the CP parities in \eqref{cp.parities}.

\begin{figure}[h]
\includegraphics[scale=0.55]{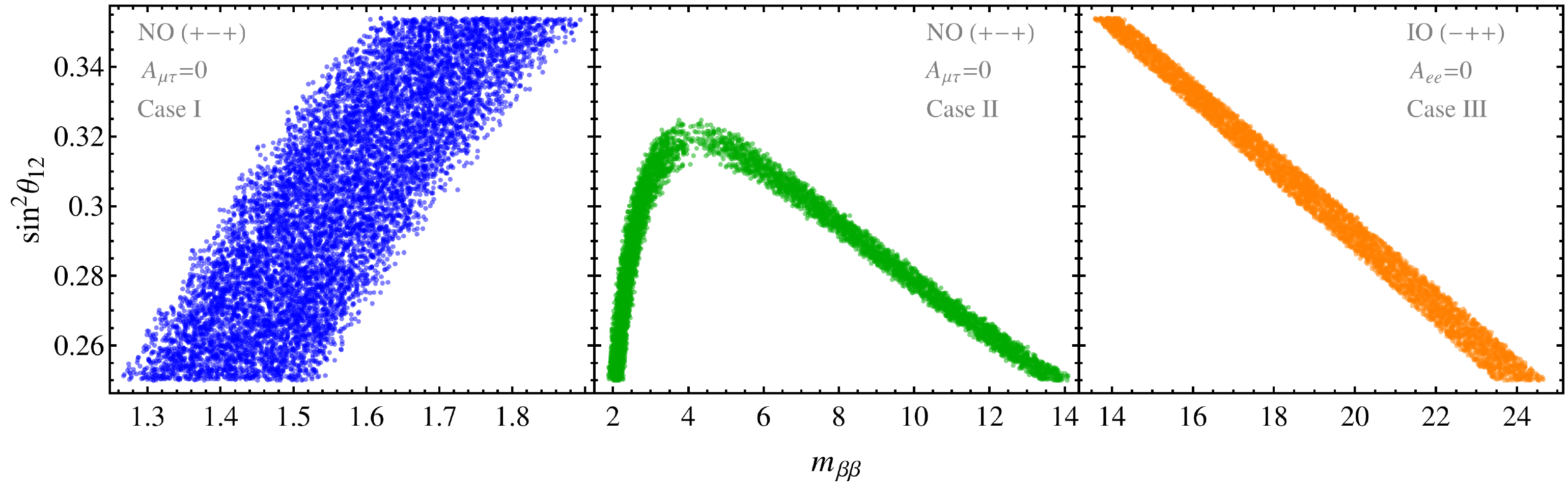}
\caption{\label{fignu.1}
Correlation between $\sin^2\theta_{12}$ and $m_{\beta\beta}$ for $\cpmutau$ symmetric neutrino mass matrix with one-zero textures in $M_\nu^{-1}$.
The oscillation observables are varied within 3-$\sigma$ of Ref.\,\cite{capozzi.17}
and $A_{\alpha\beta}$ denote $(M_\nu^{-1})_{\alpha\beta}$.
}
\end{figure}

\begin{figure}[h]
\includegraphics[scale=0.55]{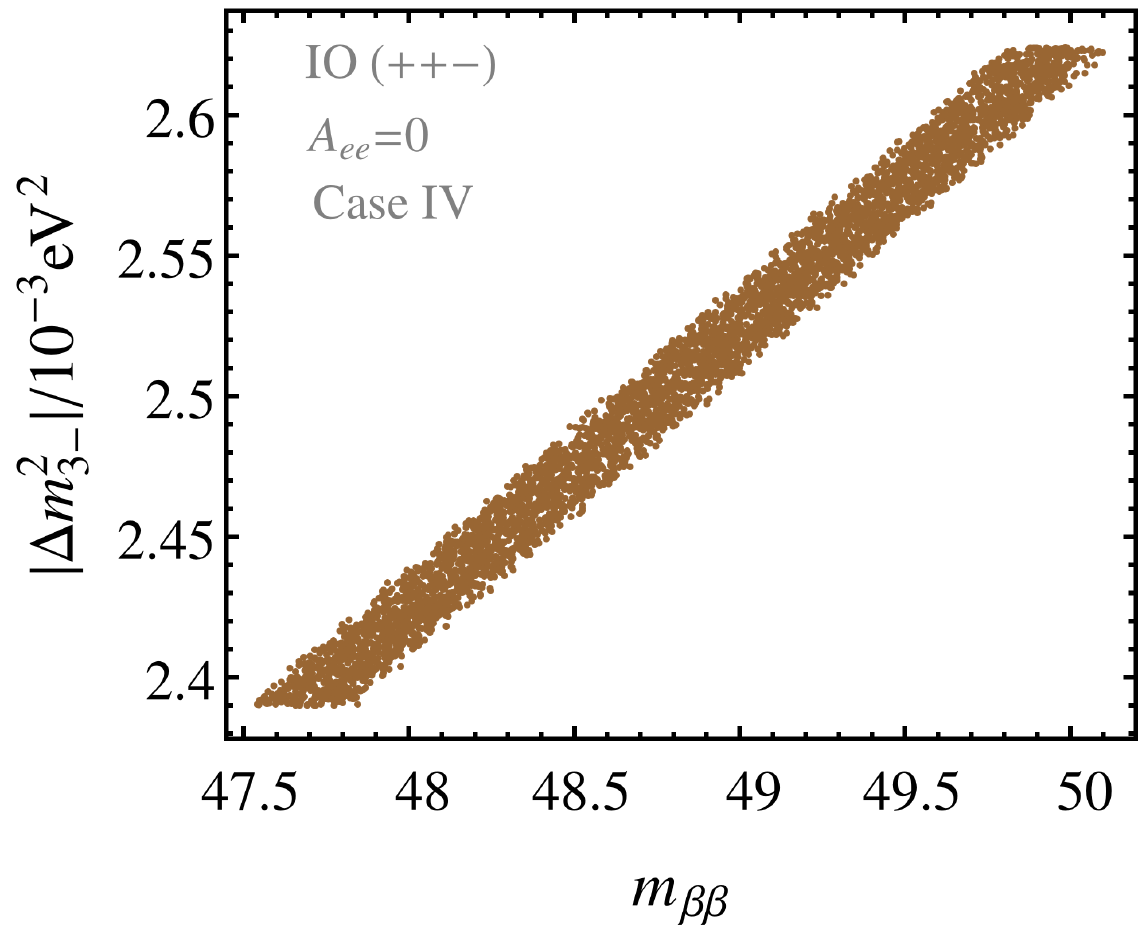}
\caption{
\label{fignu.2}
Correlation between $|\Delta m^2_{3-}|=\big|m_3^2-(m^2_1+m^2_2)/2\big|$ and $m_{\beta\beta}$ for $\cpmutau$ symmetric 
neutrino mass matrix with one-zero texture in $A_{ee}=(M_\nu^{-1})_{ee}$ (case IV).
The oscillation observables are varied within 3-$\sigma$ of Ref.\,\cite{capozzi.17}.
}
\end{figure}

As a further prediction of our scenario, various correlations between measured and unmeasured observables are expected due to the reduced number of parameters.
We show in Fig.\,\ref{fignu.1}, for cases I, II and III in table \ref{table.results}, the correlation between $\sin^2\theta_{12}$ and the yet to be measured effective parameter
\eq{
m_{\beta\beta}=|(M_\nu)_{ee}|=\left|\sum m'_i{\Uz_{ei}}^2\right|\,,
}
which controls the neutrinoless double beta decay ($0\nu\beta\beta$) rates induced by light neutrino exchange.
For case IV, such a correlation is weak and we show in Fig.\,\ref{fignu.2} the correlation between
$m_{\beta\beta}$ and $|\Delta m^2_{3-}|=\big|m_3^2-(m^2_1+m^2_2)/2\big|$.
It is clear that a better measurement of $\sin^2\theta_{12}$ ($|\Delta m^2_{3-}|$) will lead to a sharper prediction of $m_{\beta\beta}$ for cases I, II and III (case IV).
In special, for case II, it is predicted that $\sin^2\theta_{12}\lesssim 0.325$
and for case IV, $m_{\beta\beta}$ is within reach of the future experiments such as KamLAND-Zen.

\section{Heavy neutrinos}
\label{sec:heavy}

Here we show the spectrum and the mixing pattern of heavy neutrinos.
We denote the states with definite masses by $\cN_{i}$, $i=1,2,3$.
All parameters of the mass matrix for light neutrinos were determined in 
Sec.\,\ref{sec:nu} and the discrete possibilities were listed in 
table~\ref{table.results}.
Then all the information on the heavy neutrino mass matrix follows from \eqref{MR}.
There are only two free parameters: $m_D$ (or $\yN{e}$) and $\kappa$.
The first will set the overall scale for the heavy neutrino masses $M_i$, $i=1,2,3$, while $\kappa$ will determine the mass ratios and mixing pattern.
Note that we will not follow the usual convention where $(\cN_1,\cN_2,\cN_3)$ are ordered from lighter to heavier states and then it is useful to denote the lightest heavy neutrino as $\cN_0$ and its mass as $M_0$.

We can continue the analysis of the case of $\kappa=1$, which we started in 
Sec.\,\ref{sec:model}.
In this case, Eq.\,\eqref{MR} implies that the heavy neutrino mass matrix is proportional to the inverse of the light neutrino mass matrix and the diagonalizing matrix is completely fixed by the PMNS matrix; cf.\,\eqref{UR:k=1}.
The values of the heavy masses are completely determined by \eqref{Mi:mi:k=1}, except for an overall scale.
From the solar mass splitting we always have $M_2<M_1$ and the ratio is fixed by
\eq{
\label{M1/M2:k=1}
\frac{M_1^2}{M_2^2}=\frac{m_2^2}{m_1^2}=1+\frac{\Delta m^2_{21}}{m_1^2}
\approx 1+ \left(\frac{8.6\,\unit{meV}}{m_1}\right)^2\,.
}
For the NO solutions of table~\ref{table.results}, at most a mild hierarchy of $M_1/M_2\sim 3.6$ is expected.
In contrast, for IO, $m_1$ is not the lightest mass and it is more useful to rewrite
\eq{
\label{M1/M2:k=1:IO}
\frac{M_1^2}{M_2^2}=\frac{m_2^2}{m_1^2}
\approx 1+ \frac{(8.6\,\unit{meV})^2}{m_3^2+(50\,\unit{meV})^2}\,.
}
For both cases III and IV, $M_1$ is only about 1.5\% larger than $M_2$ and the 
pair $\cN_1$-$\cN_2$ is nearly degenerate.
The ordering for $M_3$, on the other hand, depends on whether the ordering follows the NO or IO:
\eqali{
\label{M1/M3::k=1}
\text{NO}:&\quad M_3<M_2<M_1\quad\text{and}\quad
\frac{M_1^2}{M_3^2}&=\frac{m_3^2}{m_1^2}=1+\frac{\Delta m^2_{31}}{m_1^2}
\approx
1+\left(\frac{50\,\unit{meV}}{m_1}\right)^2
\,,
\cr
\text{IO}:&\quad M_2<M_1<M_3\quad\text{and}\quad
\frac{M_3^2}{M_1^2}&=\frac{m_1^2}{m_3^2}=1+\frac{\Delta m^2_{13}}{m_3^2}
\approx
1+\left(\frac{50\,\unit{meV}}{m_3}\right)^2
\,.
}
From these relations, a hierarchy of at most $M_1/M_3\sim 20$  or
$M_3/M_1\sim 50$ is possible for NO or IO, respectively.
The least hierarchical case, $M_1/M_3\sim 2$, is possible for case II.
We see that the lightest mass is $M_0=M_3$ for NO and $M_0=M_2$ for IO.

The mixing matrix $U_R$ is also fixed by \eqref{UR:k=1} for $\kappa=1$.
The first row of $U_R$ should have values 
\eq{
\label{URei:k=1}
|U_{Re1}|\sim 0.83,
\quad
|U_{Re2}|\sim 0.54,
\quad
|U_{Re3}|\sim 0.15.
}
The CP parities of the heavy neutrinos are also fixed by the relation \eqref{Mi:mi:k=1}: they are opposite to the CP parities of light neutrinos, i.e.,
\eq{
\label{cp.parities:k=1}
{-K_R^2}=K_\nu^2\,.
}

When $\kappa$ deviates away from unity, the mass spectrum will cease to obey Eqs.\,\eqref{M1/M2:k=1} or \eqref{M1/M3::k=1} and $U_R$ will no longer obey \eqref{UR:k=1}.
Nevertheless, we can still establish that $-K_R^2$ and $K_\nu^2$ should have the same signature, i.e., they are the same except for possible permutations.
The proof is shown in appendix \ref{ap:KR:Knu}.
The result is that a clever choice of ordering for $M_i$ allows us to maintain \eqref{cp.parities:k=1}.
A possibility is to order the heavy neutrinos in such a way that \eqref{Mi:mi:k=1} is valid when we continually take the limit%
\,\footnote{%
In practice \eqref{MR:U} isolates the eigenvalue $M_i'$ that have the unique CP 
parity [$-(K_R^2)_{ii}<0$] because the massless case never occurs.
The remaining $M'_i$ of the same sign never cross and they can be tracked unambiguously; see discussion around \eqref{C*D2<>0}.
\label{footnote.1}
}
to $\kappa=1$.
In the same limit, $\Uz_R$ should approach $\Uz_\nu$.
With this ordering convention, we can extend the possible CP parities in Eq.\,\eqref{cp.parities} to the heavy neutrinos:
\eq{
\label{cp.parities:nu.N}
{-K_R^2}=K^2_\nu:\quad
\{(+++)\} \text{ ~or one of~ } \big\{(-++),(+-+),(++-)\big\}\,.
}
Obviously, only the second set is allowed for texture-zero solutions in Table\,\ref{table.results}.

\begin{figure}[h]
\includegraphics[scale=0.7]{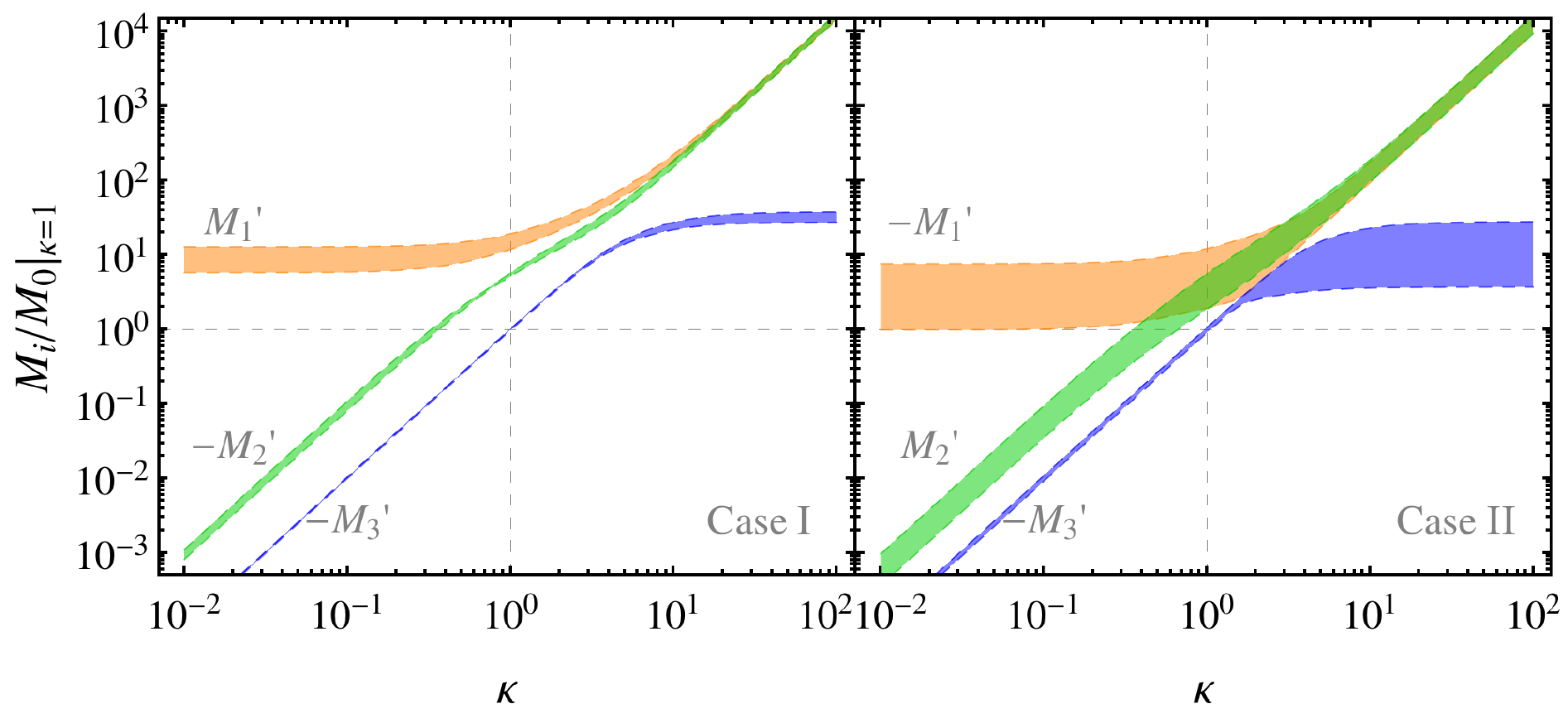}
\caption{\label{figMi.NO}
Mass spectrum for NO solutions (cases I and II) relative to the lightest mass $M_0$ for $\kappa=1$, for $\cN_1$ (orange), $\cN_2$ (green) and $\cN_3$ (blue), as a function of $\kappa$.
We use the 3-$\sigma$ ranges in Ref.\,\cite{capozzi.17} for the observables not fixed by symmetry.
$M_i'$ indicate the heavy neutrino masses with their CP parity.
}
\end{figure}

\begin{figure}[h]
\includegraphics[scale=0.7]{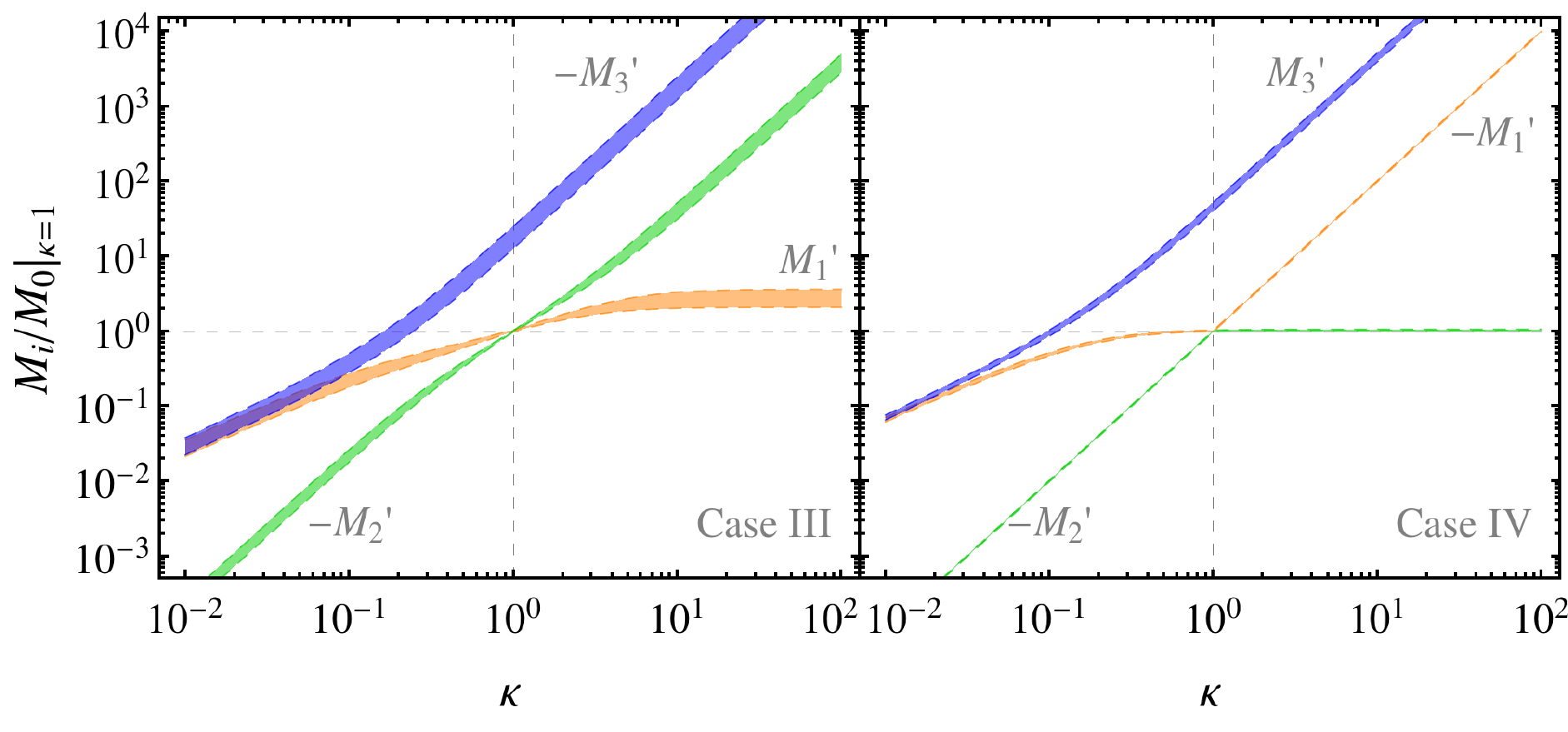}
\caption{\label{figMi.IO}
Mass spectrum for IO solutions (cases III and IV) relative to the lightest mass $M_0$ for $\kappa=1$, for $\cN_1$ (orange), $\cN_2$ (green) and $\cN_3$ (blue), as a function of $\kappa$.
We use the 3-$\sigma$ ranges in Ref.\,\cite{capozzi.17} for the observables not fixed by symmetry.
$M_i'$ indicate the heavy neutrino masses with their CP parity.
}
\end{figure}

We show how the heavy neutrino spectrum depends on $\kappa$ in Fig.\,\ref{figMi.NO} for NO (cases I and II) and in Fig.\,\ref{figMi.IO} for IO (cases III and IV) by plotting the possible values for the heavy masses $M_i$ relative to the lightest mass $M_0|_{\kappa=1}$ at $\kappa=1$.
We clearly see that the mass spectrum obeys \eqref{M1/M2:k=1} [or \eqref{M1/M2:k=1:IO}]
and  \eqref{M1/M3::k=1} for $\kappa=1$. 
To make the plots, we diagonalize $M_R$ in \eqref{MR} explicitly, keeping the convention in \eqref{cp.parities:nu.N}, and vary the observables not fixed by symmetry within their 3-$\sigma$ values reported in Ref.\,\cite{capozzi.17} by random sampling. Then the minimal and maximal values are extracted to draw the borders.%
\footnote{
For case II in Fig.\,\ref{figMi.NO}, there are regions inside the wide bands with 
very low point density, exactly in the region where the two distinct solutions 
intersect. 
}
We also indicate the CP parities for each $\cN_i$ and we see that the convention in \eqref{cp.parities:nu.N} is enough to separate $M_1$ from $M_2$ for both cases II and III.
For case IV, it seems that $M_1$ and $M_2$ cross near $\kappa=1$ but one can check by varying only $\kappa$ that they never cross.
The minimal value of $|M_i-M_0|$ for this case is checked to be 1.2\% of $M_0=M_2$.
An alternative way to gain analytic information of the heavy masses from the light neutrino masses are shown in appendix \ref{ap:Mi}.

We can prove generically that when their CP parities are included 
no crossing of eigenvalues $M_i'$ occurs when $\kappa$ is continuously changed.
The proof utilizes the rephasing invariant in \eqref{c*d2} adapted to $M_R$ when parametrized as \eqref{MR:param}:
\eq{
\label{C*D2}
\im(C^*D^2)=\ums{2}S_{13}C^2_{13}S_{12}C_{12}(M_1'-M_2')(M_2'-M_3')(M_3'-M_1')\,.
}
The diagonalizing matrix ${\Uz_R}^*$ is parametrized as \eqref{U0:param} after 
appropriate rephasing of the second and third rows, and the respective angles are 
replaced as $\theta_{ij}\to \Theta_{ij}$ with upper case $C_{ij},S_{ij}$ denoting e.g.\ 
$C_{ij}=\cos\Theta_{ij}$.%
\footnote{We use the convention that $\Uz_R$ diagonalizes $M_R^*$ and not $M_R$.
The equality \eqref{UR:k=1} implies that $\Theta_{ij}=\theta_{ij}$ for $\kappa=1$.
}
Then the relation \eqref{MR} allows us to conclude that
\eq{
\label{C*D2<>0}
\im(C^*D^2)=-m_D^6\kappa^4\im(c^*d^2)\neq 0\,,
}
i.e., it never vanishes due to \eqref{c*d2}.
Hence $M'_i$ never cross.

\begin{figure}[h]
\includegraphics[scale=0.7]{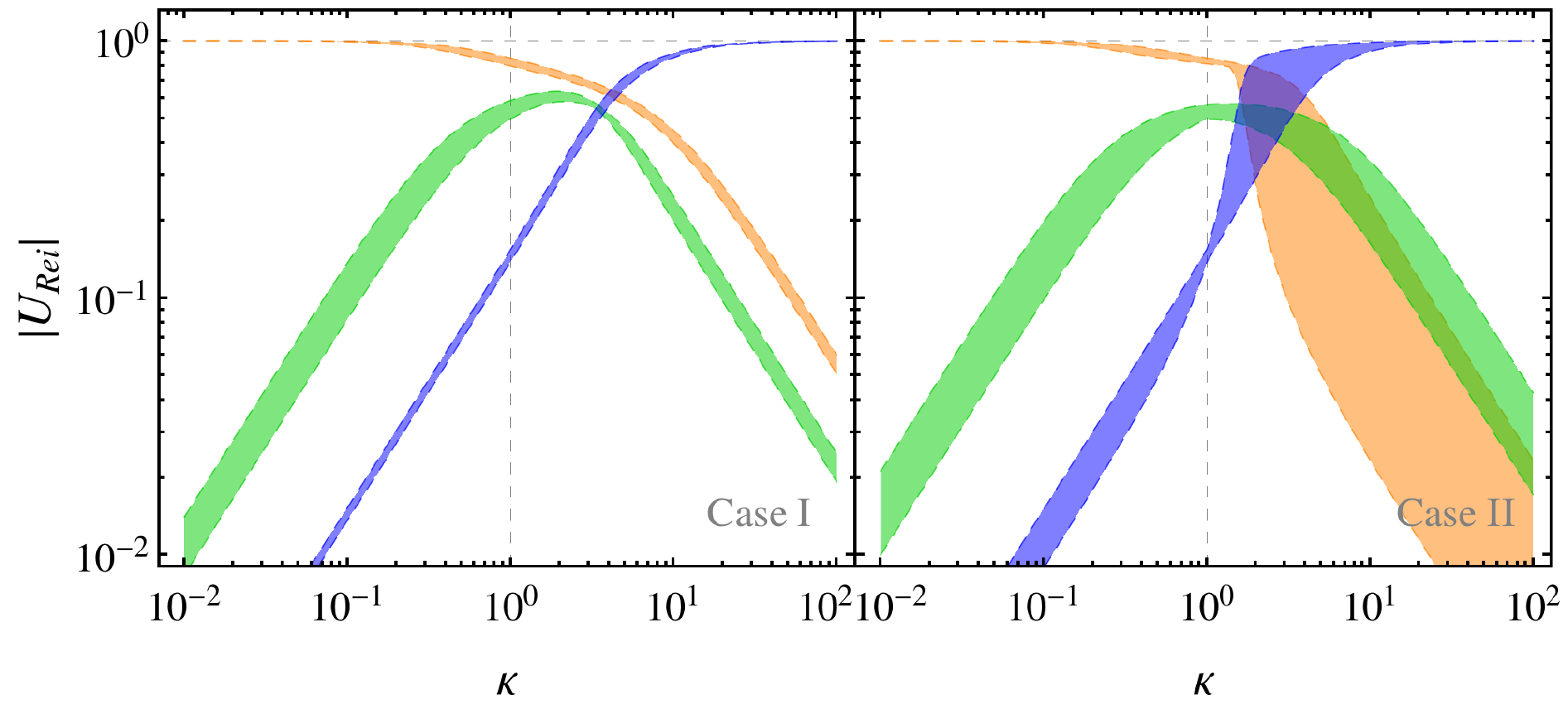}
\caption{\label{figUR.NO}
Modulus of $U_{R1i}$ for our NO solutions (cases I and II) for $i=1,2,3$ (orange, green, blue) as a function of $\kappa$.
We use the 3-$\sigma$ ranges in Ref.\,\cite{capozzi.17} for the observables not fixed by symmetry.
}
\end{figure}

\begin{figure}[h]
\includegraphics[scale=0.7]{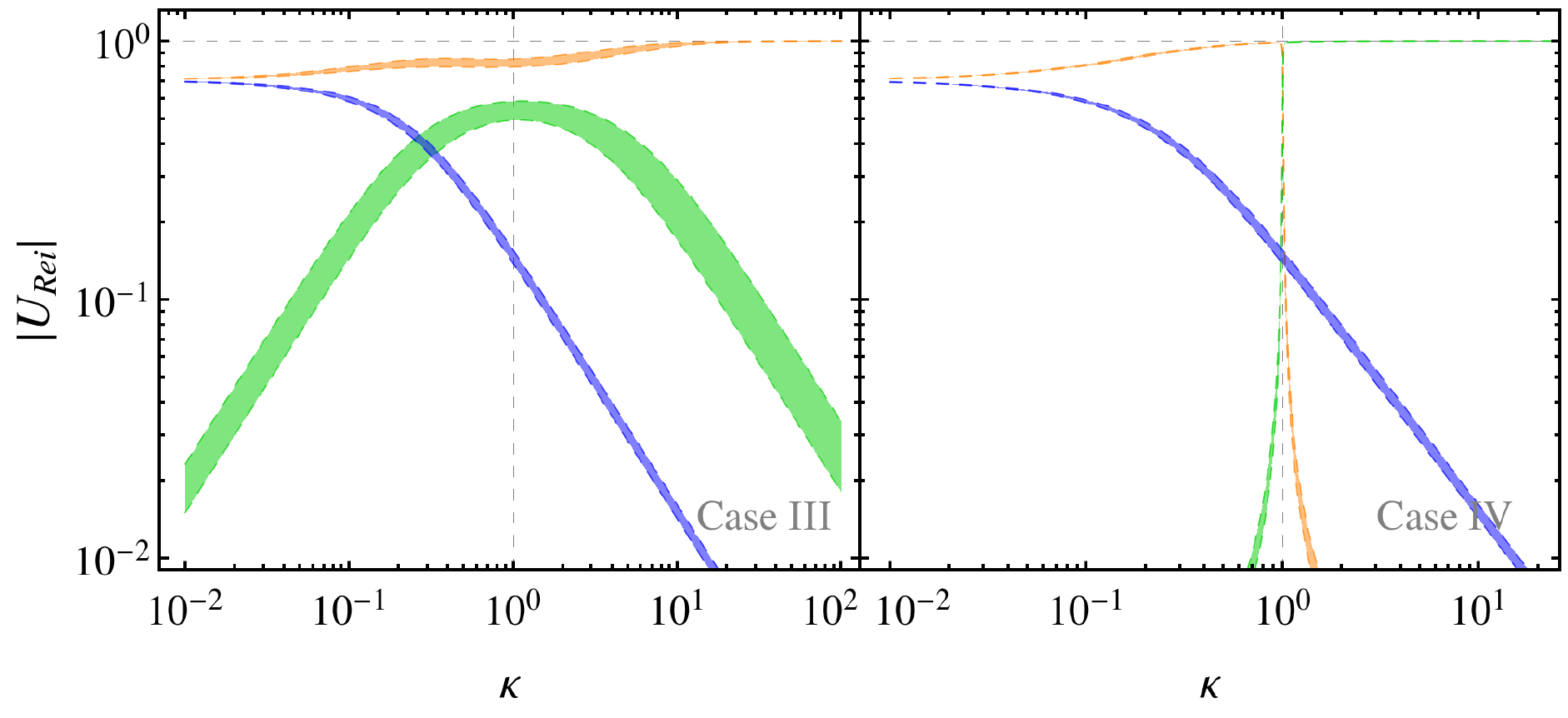}
\caption{\label{figUR.IO}
Modulus of $U_{R1i}$ for our IO solutions (cases III and IV) for $i=1,2,3$ (orange, green, blue) as a function of $\kappa$.
We use the 3-$\sigma$ ranges in Ref.\,\cite{capozzi.17} for the observables not fixed by symmetry.
}
\end{figure}

We can now turn to the mixing matrix $U_R$.
To show how the mixing matrix $U_R$ deviates from $i U_\nu$ for $\kappa\neq 1$, we need a parametrization for $U_R$.
We use the decomposition in \eqref{U=U0.K} and the parametrization in \eqref{U0}.
Two among the three entries $u_i=|U_{R1i}|$ in the first row are enough to recover the entire matrix $\Uz_R$\,\cite{mutau-r:HS}.
The procedure is reviewed in appendix \ref{ap:UR}.
Their behavior can be seen in Fig.\,\ref{figUR.NO} for the NO cases and in 
Fig.\,\ref{figUR.IO} for the IO cases.
The limit for $\kappa=1$ is clearly in accordance with \eqref{URei:k=1} except for case IV where the rapid variation for $\kappa$ near unity makes it hard to ascertain the value of $|U_{Re1}|$ and $|U_{Re2}|$ at the exact point.
We have checked that they agree with \eqref{URei:k=1}.


\section{Leptogenesis}
\label{sec:lepto}

The SM cannot explain the present baryon asymmetry of the Universe expressed in the present abundance\,\cite{planck}:
\eq{
Y_{\Delta B}\equiv\frac{n_B-n_{\bar{B}}}{s}\bigg|_0=(8.65\pm 0.09)\times 10^{-11}\,,
}
where $n_B$ is the baryon number density and $s$ is the entropy density.
When the SM is extended through some form of seesaw mechanism to account for naturally small neutrino masses, leptogenesis arises as a natural mechanism to explain the baryon asymmetry\,\cite{yanagida}.
In the simplest type I seesaw mechanism, a lepton number asymmetry is generated when the lightest heavy Majorana neutrino typically decays more to antileptons than leptons due to CP violating Yukawa couplings.
This lepton number asymmetry is then converted, within the SM, to a baryon asymmetry by spharelon processes that violate $\cB+\cL$ but conserve $\cB-\cL$\,\cite{sphaleron}.

The CP asymmetries in the decays of $\cN_i$ depend on the Yukawa couplings $\lambda_{i\alpha}$ that control the strength of the Yukwawa interactions $\bar{N}_i\tphi^\dag L_\alpha$, in the basis where $M_R$ is diagonal.
In our model, we simply have
\eq{
\label{lambda:k}
\lambda=\yN{e} U_R^\dag\diag(1,\kappa,\kappa)\,,
}
where $N_R=U_R \cN_R$ in our convention and $\yN{e}$ can be used insted of $m_D$.
Due to the highly constrained nature of our setting, only two free parameters govern the heavy neutrino sector. We follow the ordering convention from the $\kappa=1$ limit
and recall that the lightest heavy neutrino is denoted by $\cN_0$ and its mass by $M_0$.

The two free parameters, $\yN{e}$ and $\kappa$, cannot vary completely without limit as 
perturbativity of Yukawa couplings requires roughly that
\eq{
\label{y:perturb}
\yN{e},\kappa\,\yN{e}\lesssim \sqrt{4\pi}\,.
}
This requirement typically furnishes lower and upper values for $\kappa$. For example, if the lightest heavy neutrino mass is $M_0=10^{12}\,\unit{GeV}$, we will be restricted to $10^{-2}\lesssim \kappa\lesssim 10^2$.
For lower $M_0$, the allowed range increases proportionally to $M_0^{-1/2}$.
See Eq.\,\eqref{yNe=} in the following.

In the context of $\cpmutau$ symmetric models, it is known for some time that leptogenesis induced by singlet heavy neutrinos cannot proceed in the one-flavor regime where $T\sim M_0\gtrsim 10^{12}\,\unit{GeV}$\,\cite{mutau-r:GL}; see also Ref.\,\cite{cp.mutau}.
The reason is that $\cpmutau$ restricts the flavored CP asymmetries $\epsz_{\alpha}$ in the decay $\cN_0\to L_\alpha+\phi$ to obey\,\cite{cp.mutau}
\eq{
\label{cp:eps}
\epsz_{e}=0\,,~~\eps_{\mu}^{(0)}=-\eps_{\tau}^{(0)}\,.
}
Hence, the total CP asymmetry vanishes,
\eq{
\epsz=\epsz_{e}+\epsz_{\mu}+\epsz_{\tau}=0\,,
}
and a net lepton number asymmetry cannot be generated.
Only in the flavored regime\,\cite{flavored,flavored.efficiency} where the $\tau$ flavor can be distinguished by fast Yukawa interactions, i.e., when
$10^{9}\,\unit{GeV} \lesssim T\sim M_0\lesssim 10^{12}\,\unit{GeV}$, leptogenesis can be successful in generating enough lepton number asymmetry\,\cite{cp.mutau}.
See Ref.\,\cite{spectator} for a recent analysis of the temperature regimes where the various SM interactions enter in equilibrium.
Below $10^{9}\,\unit{GeV}$, where all lepton flavors can be distinguished,
Ref.\,\cite{cp.mutau} concluded within analytical approximations that leptogenesis cannot proceed because the washout in the $\mu$ and $\tau$ flavors are equal, so that the asymmetries \eqref{cp:eps} in these flavors are summed to zero.
So our case is a particular case of purely flavored leptogenesis\,\cite{purely} with the distinction that the vanishing of $\eps^{(0)}$ is protected by $\cpmutau$ and not by $\mathtt{B-L}$.
It is also a particular case, enforced by symmetry, of a case where the baryon asymmetry is generated only by the low energy Dirac CP phase and no CP violation is present in the heavy neutrino sector\,\cite{lepto:low}.

The equality of the washout effects for $\mu$ and $\tau$ flavors follows because,
in the approximation where off-shell $\Delta L=2$ scatterings and off-diagonal correlations through the $A$-matrix are neglected, 
these washout effects are controlled by the three washout parameters
\eq{
\label{m-tilde}
\tm_{0\alpha}=|\lambda_{0\alpha}|^2\frac{v^2}{M_0}\,,
\quad \alpha=e,\mu,\tau,
}
where $v=174\rm GeV$ in the SM and the subscript $0$ refers to $\cN_0$.
With $\cpmutau$ symmetry, 
\eq{
\tm_{0\mu}=\tm_{0\tau}\,,
}
and the strength of washout is the same in the latter flavors\,\cite{cp.mutau}.
In our model, this fact can be directly checked for \eqref{lambda:k}.
Current neutrino parameters implies that typically $\tilde{m}_{0}=\sum_\alpha \tilde{m}_{0\alpha}\gg m_*\approx 1.07\,\text{meV}$ and $\cN_0$ reaches the equilibrium density rather quickly and a strong washout of lepton flavors takes place depending on $\tilde{m}_{0\alpha}\gg m_*$.
The mass $m_*\equiv \frac{16\pi^2v_u^2}{3\Mp }\sqrt{\frac{g_*\pi}{5}}$ quantifies the expansion rate of the Universe.

So we focus on the intermediate regime where $10^9\lesssim T\sim M_0\lesssim 10^{12}\,\unit{GeV}$ and neglect the possible asymmetries generated by the decay of heavier $\cN_i$. We comment on possible effects in the end.
In this regime, only the $\tau$ Yukawa interactions are in equilibrium and then only the $\tau$ flavor and its orthogonal
combination are resolved by interactions. In this case, the final baryon asymmetry can be approximated by\,\cite{nardi}
\eq{
  \label{YB:2f}
Y_{\Delta B}\simeq -\frac{28}{79}\times Y^{\rm eq}_{\cN_0}\times\bigg[
\eps_2^{(0)}\eta\Big(\frac{417}{589}\tilde{m}_{02}\Big)
+\eps_\tau^{(0)} \eta\Big(\frac{390}{589}\tilde{m}_{0\tau}\Big)
\bigg]\,,
}
where $\eps_2^{(0)}=\eps_e^{(0)}+\eps_\mu^{(0)}$, $\tilde{m}_{02}=\tilde{m}_{0e}+\tilde{m}_{0\mu}$,
and the efficiency factor
\eq{
\label{eta:ap}
\eta(\tilde{m})\simeq \bigg(
\Big(\frac{\tilde{m}}{2.1m_*}\Big)^{-1}
+\Big(\frac{2\tilde{m}}{m_*}\Big)^{1.16}
\bigg)^{-1}\,,
}
is valid for the strong washout regime but allows weak or mild washout in some flavor\,\cite{flavored.efficiency}.
The factors 417/589 and 390/589 correspond to the diagonal entries of the $A$
matrix and quantifies the effects of flavor in the washout processes when changing
from the asymmetry in lepton doublets to asymmetries in $\Delta_\alpha=\cB/3-\cL_\alpha$\,\cite{flavored.efficiency}; see also \cite{nardi}.
We ignore the small effects of off-diagonal elements of the $A$ matrix
and consider the third family Yukawas in equilibrium as well as $h_c$.
We can see that the properties \eqref{cp:eps} of $\cpmutau$ leads to a partial
cancellation of the baryon asymmetry in \eqref{YB:2f} but it is nonzero because
the $\tau$ flavor and its orthogonal combination are washed out differently.
The quantity $Y^{\rm eq}_{\cN_0}$ is the equilibrium thermal density of $\cN_0$ per total entropy density and is given by
$
Y^{\rm eq}_{\cN_0}=
\frac{135\zeta(3)}{4\pi^4g_*}\approx 3.9\times 10^{-3}\,,
$
where the last numerical value is for the SM degrees of freedom below the $\cN_0$
mass ($g_*=106.75$).
The factor $28/79$ corresponds to the reduction of the asymmetry in
$\Delta_\alpha$ to $\cB-\cL$ in the SM due to 
spharelons when they go out of equilibrium before EWPT.

In the $\cpmutau$ symmetric case, we can rewrite \eqref{YB:2f} in the form
\eq{
\label{YB:2f:mutau}
Y_{\Delta B}= -\frac{28}{79}\times Y^{\rm eq}_{\cN_0}\times
\eps^{(0)}_\tau\big(\eta_{0\tau}-\eta_{02})\,,
}
where we denote
\eq{
\eta_{0\tau}=
	\eta\Big(\frac{390}{589}\tilde{m}_{0\tau}\Big)
\,,\quad
\eta_{02}=
	\eta\Big(\frac{417}{589}\tilde{m}_{02}\Big)
\,.
}
One can note that the sign of the final baryon abundance is determined by the sign of $-\eps^{(0)}_\tau$ because the combination $\eta_\tau-\eta_2>0$, as the washout function \eqref{eta:ap} is a decreasing function in the strong washout regime where $\tilde{m}_{0} > m_*$.

The necessary CP asymmetry in the $\tau$ flavor, in the generic type I seesaw case, can be written as
\eqali{
\label{eps:lambda}
\eps^{(0)}_\tau&=-\frac{1}{8\pi(\lambda\lambda^\dag)_{00}}
\sum_{j\neq 0}\bigg\{
\im\big[(\lambda\lambda^\dag)_{j0}\lambda_{j\tau}\lambda^*_{0\tau}\big]g(x_j)
\cr
&\quad +\
\im\big[(\lambda\lambda^\dag)_{0j}\lambda_{j\tau}\lambda^*_{0\tau}\big]
\frac{1}{1-x_j}\bigg\}\,,
}
where $x_j\equiv M^2_j/M_0^2$ and
\eq{
g(x)\equiv \sqrt{x}\big[\frac{1}{1-x}+1-(1+x)\ln\big(\frac{1+x}{x}\big)\big]
\equiv \frac{\sqrt{x}}{1-x}+f(x)\,.
}
The part proportional to $f(x)$, the vertex function, corresponds to the one-loop vertex contribution while the rest corresponds to the self-energy contribution for $N_R$.
We are assuming that $\cN_{j}$ masses are hierarchical, i.e., $|M_j-M_0|\gg
\Gamma_0$ for $\cN_j$ different from the lightest one and the $\cN_0$ decay width is 
\eq{
\Gamma_0=\frac{M_0}{8\pi}(\lambda\lambda^\dag)_{00}\,.
}
It is easy to see that for $\kappa=1$, the flavored CP asymmetry \eqref{eps:lambda} is vanishing as $(\lambda\lambda^\dag)_{ij}\propto \delta_{ij}$ due to our simple form \eqref{lambda:k}.
Therefore, at least a small departure from $\kappa=1$ is necessary to obtain a nonzero abundance.
In fact, the expression in \eqref{eps:lambda} can be simplified to 
\eq{
\label{epstau:simp}
\eps^{(0)}_\tau= \frac{\yN{e}^2\kappa^2(1-\kappa^2)}{\kappa^2+(1-\kappa^2)|\Uz_{Re0}|^2}
\times(\text{function of $U_R$ and $x_j$})\,.
}
The full expression is shown in appendix \ref{ap:eps}.

We can now analyze how the different quantities depend on our free parameters $\kappa$ and $\yN{e}$.
It is clear from \eqref{MR} and \eqref{lambda:k} that $M_R$ and $\lambda_{i\alpha}$ scale as $\yN{e}^2$ and $\yN{e}$, respectively. Then mass ratios $M_i/M_0$ and
$\tilde{m}_{0\alpha}$ in \eqref{m-tilde} are independent of $\yN{e}$ and only depend on $\kappa$.
On the other hand, the CP asymmetry in \eqref{eps:lambda} scales as $\yN{e}^2$ and that is also the scaling behaviour of the baryon abundance in \eqref{YB:2f:mutau}.
Therefore, the only dependence of $Y_{\Delta B}$ on $\yN{e}$ can be factorized as $\yN{e}^2$ while the remaining expression only depends on $\kappa$.

It is much more convenient, however, to consider the lightest heavy mass $M_0$ as the free parameter instead of $\yN{e}$,  for each $\kappa$.
We can trade $\yN{e}$ for $M_0$ as follows.
First, we factor the dependence of the lightest eigenvalue of $M_R$ on $\kappa$ with fixed $\yN{e}$ by defining
\eq{
\label{f0}
f_0(\kappa)\equiv \frac{\min_i\{M_i\}}{M_0|_{\kappa=1}}\,.
}
The masses $M_i$ are calculated from the eigenvalues of \eqref{MR} with fixed $\yN{e}$, say $\yN{e}=1$.
Generically, $f_0(\kappa)$ is a monotonically increasing (hence one-to-one) function with $f_0(1)=1$ but not smooth when there is a crossing of $M_i$ (differently for $M_i'$ which never cross).
This function can be seen in the blue band of Fig.\,\ref{figMi.NO} for NO where $M_0=M_3$ for all $\kappa$.
The band is due to the variation within 3-$\sigma$ of the low energy observables not fixed by symmetry.\footnote{%
For numerical computations we use a fixed value for $M_0|_{\kappa=1}$ averaged over
the oscillation observables not fixed by symmetry.
Hence the variation on the latter observables only appears in the numerator of 
$f_0$.
This procedure explains the small finite thickness of the low lying curve even at $\kappa=1$.
}
For IO, $M_0=M_2$ or $M_0=M_3$ depending on $\kappa$ for case III and always $M_0=M_2$
for case IV. 
The function $f_0$ is shown in the low-lying green-orange (green) band of Fig.\,\ref{figMi.IO} for case III (IV).
The transition from $M_0=M_2$ to $M_0=M_3$ for case III leads to discontinuities in $\lambda_{0\alpha}$ due to reordering of $U_{R\alpha 0}$; see Fig.\,\ref{figUR.NO}. These in turn, lead to jumps in $\tilde{m}_{0\alpha}$ for this case.

As a second step, we define a reference value for $M_0$:
\eq{
\bar{M}_0\equiv 
M_0\big|_{\kappa=1,\yN{e}=1}
=\frac{v^2}{m_{\max}}
=6.05\times 10^{14}\,\unit{GeV}\times\left(\frac{50\,\unit{meV}}{m_{\max}}\right)\,,
}
where $m_{\max}	$ is the heaviest light neutrino mass: $m_3$ for NO and $m_2$ for IO.
The dependence of $M_0$ on $\kappa$ and $\yN{e}$ can be made explicit as 
\eq{
M_0=\yN{e}^2f_0(\kappa)\bar{M}_0\,.
}
The inverse relation then gives $\yN{e}$ as a function of $M_0$ for each $\kappa$:
\eqali{
\label{yNe=}
\yN{e}^2&=
\frac{M_0}{\bar{M}_0}\frac{1}{f_0(\kappa)}
\cr
&=
1.65\times 10^{-3}\times\left(\frac{M_0}{10^{12}\,\unit{GeV}}\right)
\left(\frac{m_{\max}}{50\,\unit{meV}}\right)\frac{1}{f_0(\kappa)}\,.
}
Hence, $\yN{e}$ is completely determined for each $M_0$ (scaling as $\sqrt{M_0}$) and $\kappa$. For example, the perturbativity requirement in \eqref{y:perturb} can be easily extracted.
The relation \eqref{yNe=} and the function \eqref{f0} accomplish the purpose of expressing all the relevant quantities in the baryon asymmetry \eqref{YB:2f:mutau} solely in terms of $\kappa$ and $M_0$.
Moreover, the dependence on $M_0$ is only multiplicative as
\eq{
\label{YB.M0}
Y_{\Delta B}=M_0\times (\text{function of $\kappa$}).
}

Using \eqref{f0} we can write, for example, the explicit dependence on $\kappa$ of
\eq{
\tm_{0\alpha}=\frac{|\tla_{0\alpha}|^2}{f_0(\kappa)}m_{\max}\,,
}
where $\tla$ is the Yukawa matrix with $\yN{e}$ factored out, i.e., 
\eq{
\tla=\yN{e}^{-1}\,\lambda=U_R^\dag\diag(1,\kappa,\kappa)\,.
}
We have checked that typically $\tilde{m}_{02},\tilde{m}_\tau>20\,\unit{meV}$ and strong washout in all flavors take place.
Only for case IV, $\tilde{m}_{0\tau}\sim 0.5\text{--}0.6\,\unit{meV}$ for $\kappa>1$ and
the asymmetry in the $\tau$ flavor is washed out only mildly.

The $\cN_0$ decay width can be also rewritten as
\eq{
\frac{\Gamma_0}{M_0}=\frac{1}{8\pi}\frac{M_0}{\bar{M}_0}\frac{(\tla\tla^\dag)_{00}}{f_0(\kappa)}\,.
}
This relation allows us to check that we will be typically away from the resonant regime because
\eq{
\label{width:window}
6\times 10^{-5}\le \frac{\Gamma_0}{M_0}\le 2.4\times 10^{-4}\,,
}
for $M_0=10^{12}\,\unit{GeV}$ and our four solutions in table \ref{table.results}.
Lower values of $M_0$ will give proportionally lower ratios.

\begin{figure}[h]
\includegraphics[scale=0.65]{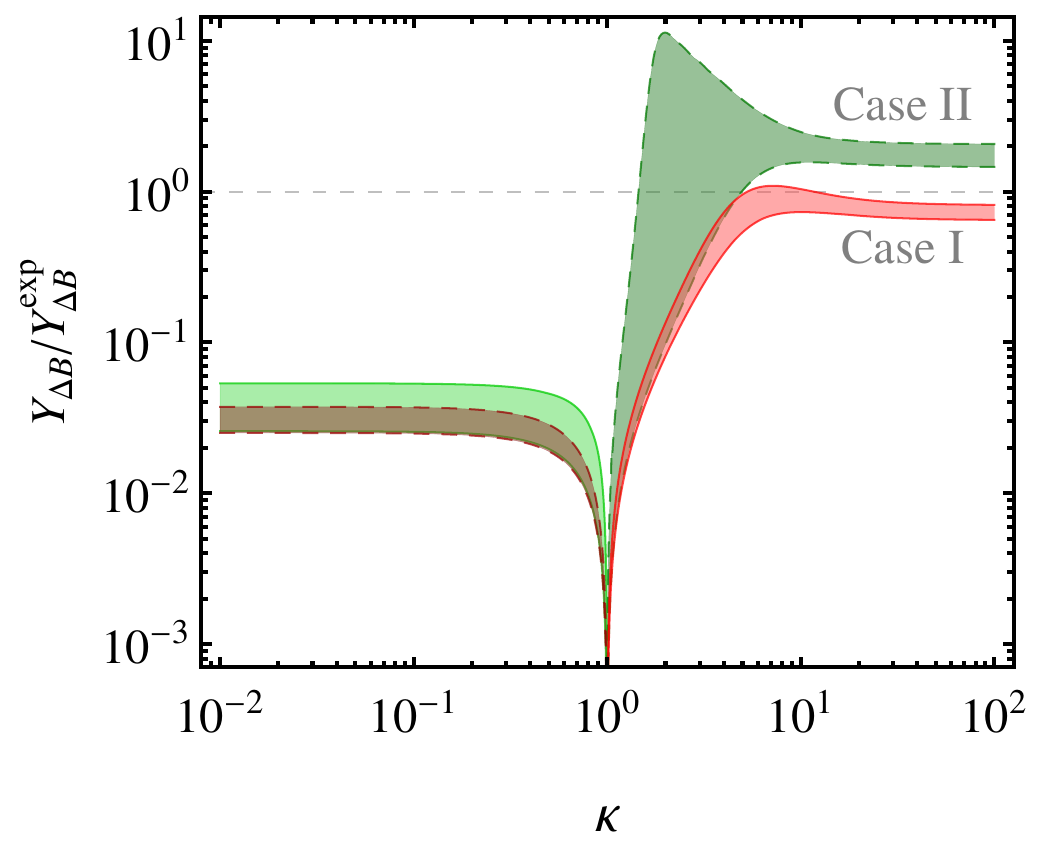}
\includegraphics[scale=0.65]{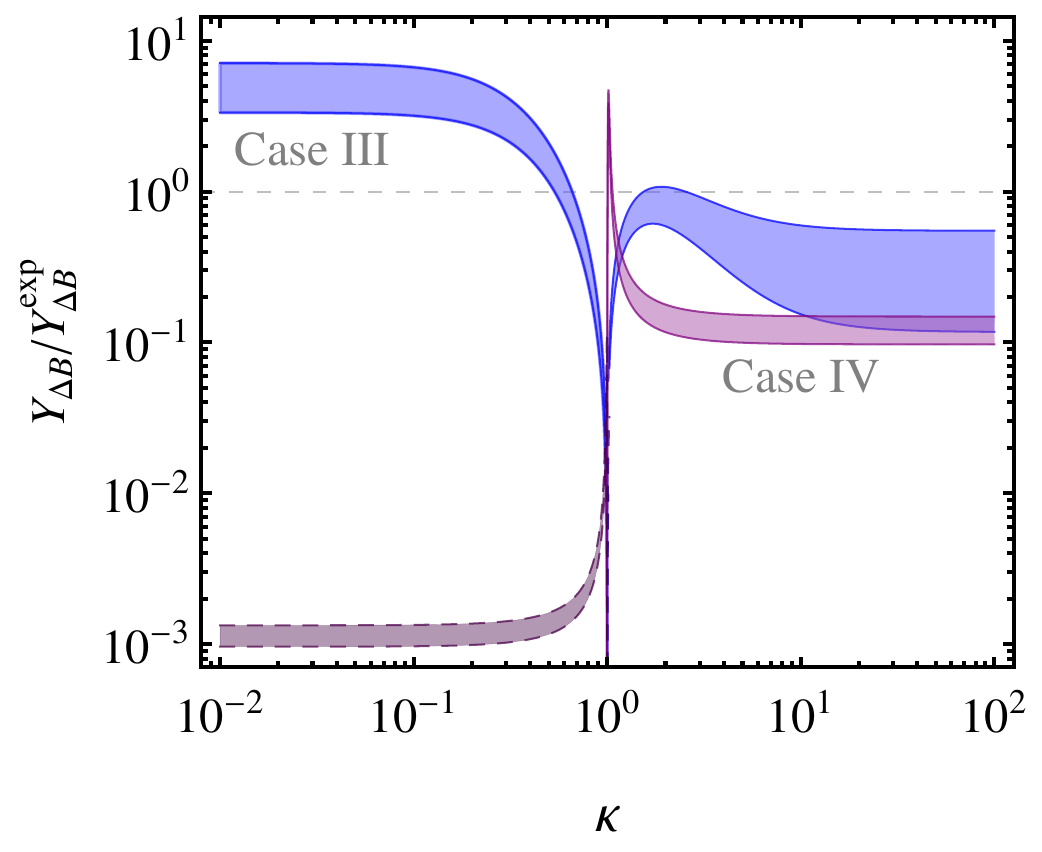}
\caption{\label{fig.YB}
Expected baryon abundance over the experimental value for the NO solutions (left) and IO solutions (right) for $M_0=10^{12}\,\unit{GeV}$ and $\delta=-90^\circ$ for the Dirac CP phase.
The regions with lighter colors and solid borders refer to positive $Y_{\Delta B}$ 
while the regions with darker colors and dashed borders refer to positive $-Y_{\Delta B}$.
The oscillation observables are varied within 3-$\sigma$ of Ref.\,\cite{capozzi.17}.
}
\end{figure}

We can now show in Fig.\,\ref{fig.YB} the baryon asymmetry $Y_{\Delta B}$ we expect for our four solutions, considering $M_0=10^{12}\,\unit{GeV}$ and $\delta=-90^\circ$ for the low-energy Dirac CP phase.
Results for lower values of $M_0$ can be reinterpreted by rescaling linearly as in
\eqref{YB.M0} down to $M_0\approx 10^9\,\unit{GeV}$ which is the lowest (approximate) value for which the flavor regime with $\tau$ resolved is still valid.
We also show $-Y_{\Delta B}$ (dashed style and darker colors) which corresponds to the baryon asymmetry for the disfavored case $\delta=90^\circ$, because flipping the sign of $\delta$ flips the signs of both $\eps_\tau^{(0)}$ and $Y_{\Delta B}$.
For the current preferred value of $\delta=-90^\circ$,
only cases I, III and IV can give the right asymmetry in certain parameter regions, some of them very narrow.
The value $\delta=+90^\circ$ is disfavored in more than 3$\sigma$ in current global fits\,\cite{capozzi.17} and case II is then the least favored.
The possible parameter regions in the $\kappa$-$M_0$ plane that can lead to 
successful leptogenesis are shown in table~\ref{table.lepto} where only the 
rectangular borders enclosing the real regions are listed.
These regions can be read off from Fig.\,\ref{fig.YB}. For example, for case I, only the region around $\kappa\approx 8$ and $M_0\approx 10^{12}\,\unit{GeV}$ survives because for a lower value of $M_0$, the red region will be scaled down proportionally and a sufficient asymmetry cannot be generated.
In all cases for $\delta=-90^\circ$, successful leptogenesis requires that $M_0$ be 
restricted to the narrow band of the intermediate region: $1.4\times 
10^{11}\,\unit{GeV}\lesssim M_0\lesssim 10^{12}\,\unit{GeV}$.
\begin{table}[h]
\begin{tabular}{|c|c|c|c|c|}
\hline
 & \multicolumn{2}{|c|}{$\delta=-90^\circ$} & \multicolumn{2}{|c|}{$\delta=+90^\circ$}
 \\
\hline
Case & $\kappa$ & $M_0/10^{12}\,\unit{GeV}$ & $\kappa$ & $M_0/10^{12}\,\unit{GeV}$ \\
\hline
I 	& 5.2 -- 12 & 0.92 -- 1 & $\times$ & $\times$ \\
II 	& $\times$ & $\times$ & 1.5 -- 100 & 0.09 -- 1 \\
III & 0.01 -- 0.66 & 0.14 -- 1  & $\times$ & $\times$ \\
	& 1.55 -- 2.57 & 0.93 -- 1  & $\times$ & $\times$ \\
IV 	& 1.004 -- 1.06 & 0.21 -- 1 & $\times$ & $\times$ \\
\hline
\end{tabular}
\caption{\label{table.lepto}
Necessary parameter range of the model for successful leptogenesis.
}
\end{table}

Few comments are in order.
Firstly, and most surprisingly, Fig.\,\ref{fig.YB} shows no divergent resonant peak for case III
where the two lightest heavy masses approach the degenerate limit near $\kappa=1$, albeit our use of the CP asymmmetry \eqref{eps:lambda} which do not include any regulator\,\cite{resonant}.
See also review in Ref.\,\cite{resonant:rev}.
The reason is that in our model the CP asymmetry $\eps^{(0)}_\tau$ do not diverge for heavy neutrino masses of opposite CP parity even in the degenerate limit because the divergence in the vertex correction is cancelled by the self-energy contribution.
See appendix \ref{ap:eps} for the explicit expression.
This feature explains the lack of divergenes in Fig.\,\ref{fig.YB} and also applies to the CP asymmetry of the heavier $\cN_i$.
For case IV, there is indeed a peak near $\kappa=1$ but there is no divergence because $M_1-M_2$ 
never really vanish.
The minimal value of $|M_i-M_0|/M_0=1.2\%$ implies that we do not reach the resonant regime and no regulator is needed since the width is much smaller; cf.\ \eqref{width:window}.

Secondly, we note that our results for successful leptogenesis listed in table 
\ref{table.lepto} should not be interpreted as precise values but rather as rough 
estimates.
The approximate formula \eqref{eta:ap} we used for the final efficiency factor has an estimated uncertainty of the order of 30\%\,\cite{flavored.efficiency}.
Some neglected effects such as thermal corrections and spectator processes may also lead to small corrections; see e.g.\,\cite{nardi}.
We also assumed that at a temperature of $10^{12}\,\unit{GeV}$ the $\tau$ Yukawa interaction is already fast enough that the $\tau$ flavor can be distinguished from the rest but, in reality, there is a transition region where some correlation among flavors may survive until $10^{11}\,\unit{GeV}$\,\cite{spectator}.
In this transition region, correlations that are off-diagonal in flavor may be important.

Another important aspect in our case is the possible effect of the heavier $\cN_i$ in the generation and washout of additional lepton asymmetry for temperatures $T>M_0$.
As can be seen in Figs.\,\ref{figMi.NO} and \ref{figMi.IO}, there are regions for the solutions for case I and case III in which the hierarchical approximation is justified.
But in other regions, the masses $M_i$ are not hierarchical and the effects of heavier $\cN_i$
may not be negligible; see Refs.\,\cite{nardi,flavored:dev,antusch} and references therein.
In fact, for all cases, there are large ranges for $\kappa$ where the ratio between the second lightest and the lightest mass is less than 10.
The mass difference may even vanish (almost vanish) for case III (IV) as discussed above.
However, as the window for successful $\cN_0$ leptogenesis is already narrowly 
restricted between $10^{11}\,\unit{GeV}$ and $10^{12}\,\unit{GeV}$, the decay of the 
heavier $\cN_i$ will not generate a lepton asymmetry if the latter is generated 
above $10^{12}\,\unit{GeV}$ where there is no flavor effect and the total asymmetry 
vanishes due to \eqref{cp:eps}, still valid in this case.
Some lepton asymmetry may be generated below $10^{12}\,\unit{GeV}$, but 
we have checked that the CP asymmetry generated by the decay of the second lightest $\cN_{SL}$ into $\tau$ flavor is at most of the same order of $\eps^{(0)}_\tau$ and the total washout parameter is large, $\tilde{m}_{SL}\gg m_*$, although the parameter for $\tau$ flavor could be smaller than unity.
So, these effects are at most of the same order and a detailed account is beyond the scope of this paper.

With the previous caveats in mind, it is worth discussing the case where the real effenciency factor is actually 10\% smaller than our approximation in \eqref{eta:ap}.
In this case, the region for case I and the second region for case III disappear 
completely, leaving only two regions of IO as viable solutions for $\delta=-90^\circ$.
Moreover, for case IV, only a narrow region near $\kappa=1$ is allowed and in this region the heavy neutrino parameters are approximately determined.
For example, the heavy masses $M_i$ are approximately proportional to $m_i^{-1}$; cf. \eqref{Mi:mi:k=1}.
In contrast, no new regions appear if the efficiency factor were 10\% larger.

\section{Conclusions}
\label{sec:concl}

We have shown a highly predictive model of leptons where the light neutrino sector is completely determined up to discrete solutions and the heavy neutrino sector responsible for the seesaw is controlled by only two free parameters.

The model implements the $\mu\tau$-reflection symmetry in the neutrino sector and its predictions of maximal atmospheric angle, maximal Dirac CP phase, and trivial Majorana phases follow.
The model allows both the maximal values $\pm 90^\circ$ for the Dirac CP phase but the negative value is currently preferred from global fits.
The predictivity is increased by additionally enforcing an abelian $\ZZ_8$ symmetry, combined nontrivially with the $\mu\tau$-reflection symmetry, that leads to one texture zero in the $(ee)$ or $(\mu\tau)$ entry of the heavy neutrino mass matrix and hence transmitted to the inverse of the light neutrino mass matrix.
No free parameters are left in the low energy theory after the neutrino observables are accommodated and only four solutions for the lightest neutrino mass are possible depending on three possible CP parity combinations.
The possible values are shown in table \ref{table.results}.
There are two solutions for normal ordering and two solutions for inverted ordering.
In all cases, except one, the lightest neutrino mass lies in the few meV range.
Only in one NO solution, the lightest mass can vary up to 30 meV.
The effective parameter that controls neutrinoless double decay through light 
neutrino exchange is completely fixed as well. One of the solutions for IO is within 
reach of the KamLAND-Zen experiment in the 800 phase which will probe the IO 
region\,\cite{kamland-zen:800}.
Due to the reduced number of parameters, correlations between the neutrinoless double beta decay parameter $m_{\beta\beta}$ and other oscillation observables arise.

In parallel, the two free parameters of the heavy neutrino sector completely control the mass spectrum and the mixing relative to the charged leptons.
One parameter sets the overall mass scale and the other controls the mass hierarchy and mixing angles.
The heavy neutrino sector is then further constrained from the requirement of 
successful leptogenesis.
Only small regions in the space of the two free parameters are allowed.
These regions can be seen in table \ref{table.lepto}.
For the preferred value of $\delta=-90^\circ$, only three out of the four solutions, one NO and two IO, allow the production of enough baryon asymmetry.
In all cases, the lightest heavy neutrino mass needs to lie roughly in the small window of $10^{11}$ to $10^{12}\,\unit{GeV}$ where flavor effects are crucial.
Since the window is narrow, the maximal amount of generated baryon asymmetry is  sensitive to the efficiency factor that quantifies the washout effects and even a 10\% reduction would eliminate the NO solution and only two small regions for IO solutions would remain.
Moreover, in our model, a resonant enhancement of the CP asymmetry is not possible if the degenerate heavy neutrinos have opposite CP parities and all our CP asymmetries are finite even without the inclusion of a regulator.

In summary, a highly predictive model of leptons is presented where all parameters of the theory, except two, are completely fixed.
These two parameters in turn controls the heavy neutrino sector and are further 
constrained to small regions from successful leptogenesis.

\acknowledgements

C.C.N.\ acknowledges partial support by brazilian Fapesp, grants 2014/19164-6 and 2018/07903-0, and
CNPq, grant 308578/2016-3.
B.L.S.V.\ is thankful for the support of Fapesp funding grant 2014/19164-6.
G.S.S.\ thanks CAPES for financial support.

\appendix
\section{$\mu\tau$ mass difference}
\label{ap:mutau}

The large mass difference between $\mu$ and $\tau$ should be generated by the breaking of $\cpmutau$ at a high scale.
As explained in \cite{cp.mutau}, that can be achieved by the vev of a CP odd scalar $\sigma_-$.
For simplicity we can assume that the CP breaking scale $\aver{\sigma_-}$ is of the same order of magnitude as the $\ZZ_8$ breaking scale which will roughly set the mass scale for the heavy right-handed neutrinos.
In our case, we need the latter to be around $10^{12}$~GeV for flavored leptogenesis to be effective.
One concrete possibility was described in Ref.\,\cite{cp.mutau} and involves the introduction of both CP even and CP odd scalars $\sigma_{\pm}$ which are odd, together with the right-handed charged leptons, under a new $\ZZ_2$ as
\eq{
  \label{Z2}
\ZZ_2:\quad \sigma_{\pm}, l_{iR}~\text{ are odd};
}
the rest of fields are even.
As both $\sigma_{\pm}$ are invariant under $\ZZ_8$ we can write an effective Lagrangian 
below a scale $\Lambda_\cp\gg \aver{\sigma_-}$ as
\eq{
  \label{lag:l:eff}
-\lag^l_{\rm eff}=\frac{\sigma_e}{\Lambda_{\cp}}\bar{L}_eHl_e
+\frac{\sigma_\mu}{\Lambda_{\cp}}\bar{L}_\mu Hl_\mu
+\frac{\sigma_\tau}{\Lambda_{\cp}}\bar{L}_\tau Hl_\tau
+h.c.
}
The quantities $\sigma_\alpha$, $\alpha=e,\mu,\tau$ are certain linear combinations 
of $\sigma_\pm$\,\cite{cp.mutau} and generate the hierarchical Yukawa couplings in 
\eqref{lag:l} after $\sigma_\pm$ acquire vevs.
An explicit UV completion can be constructed with the introduction of three vector-like charged leptons $E_i$\,\cite{cp.mutau}.

We can see that the $\ZZ_2$ above protects the neutrino sector from CP breaking effects.
The fields $\sigma_\pm$ cannot couple directly to $N_{\alpha R}N_{\beta R}$ due to such a $\ZZ_2$ and also to $\ZZ_4^{B-L}$.
The $\cpmutau$ preserving vevs in eq.\,\eqref{eta:cpc} are also not disrupted in the scalar potential because there is no direct coupling between $\sigma_-$ and some CP odd combination of $\eta_{1,3}$ because such a combination is only possible at the quartic level and no renormalizable term can be written; see the potential for $\eta_k$ in Ref.\,\cite{cpmutau:0}.

\section{CP parities for heavy neutrinos}
\label{ap:KR:Knu}

For generic $\kappa$, we can still establish that $-K_R^2$ and $K_\nu^2$ are the same, except for possible permutations.
We can show this by changing basis
\eq{
L_\alpha \to (U_{\mu\tau})_{\alpha i}L_i\,,
\quad
N^{c}_{\alpha R}\to (U_{\mu\tau})^*_{\alpha i}N^c_{iR}\,,
}
where
\eq{
\label{U:mutau}
U_{\mu\tau}\equiv
\left(
\begin{array}{ccc}
 1 & 0 & 0 \\
 0 & \frac{1}{\sqrt{2}} & -\frac{i}{\sqrt{2}} \\
 0 & \frac{1}{\sqrt{2}} & \frac{i}{\sqrt{2}} \\
\end{array}
\right)
\,.
}
Then the mass matrices are transformed to
\eqali{
\label{real.basis}
M_{\nu}\to \bM_{\nu}&=U_{\mu\tau}^\tp M_\nu U_{\mu\tau}
\,\cr
M_{R}\to \bM_{R}&=U_{\mu\tau}^\dag M_R U^*_{\mu\tau}\,,
}
where both barred matrices are real symmetric\,\cite{cp.mutau}. Hence they can be 
diagonalized by real orthogonal matrices and the real eigenvalues will have signs 
determined by $K_\nu^2$ and $K_R^2$, respectively.
Since $U_{\mu\tau}$ commutes with the Dirac mass matrix $M_D\sim \diag(1,\kappa,\kappa)$, the relation between $\bM_\nu$ and $\bM_R$ will be still analogous to \eqref{MR},
\eq{
\label{M-bar:nuR}
\bM_{R}=-m_D^2\diag(1,\kappa,\kappa)\bM_\nu^{-1}\diag(1,\kappa,\kappa)\,.
}
Then Sylvester's law tells us that $\bM_\nu^{-1}$ and $-\bM_R$ should have the same 
signature, i.e., $-K_R^2$ and $K_\nu^2$ should have the same number of positive and 
negative signs.
This result proves the possible CP parities in \eqref{cp.parities:nu.N}.

\section{Heavy neutrino masses}
\label{ap:Mi}

The spectrum of heavy neutrinos can be determined from the relation \eqref{MR} between 
light and heavy neutrino mass matrices and the diagonalization relations 
\eqref{Mnu.MR:U}.
Everything follows from the relation \eqref{M-bar:nuR} in the basis \eqref{real.basis}.
The equality between the determinants leads to the simple relation
\eq{
M_1'M_2'M_3' = -\frac{\kappa^4 m_D^6}{m_1'm_2'm_3'}\,.
}
This relation assures us that heavy masses are always finite.

The trace, for our texture-zero cases of $a=0$ or $b=0$, leads respectively to
\eq{
M_1'+M_2'+M_3' = -
\bigg(\frac{m_D^2}{m_1'}+\frac{m_D^2}{m_2'}+\frac{m_D^2}{m_3'}\bigg)
\times\left\{\begin{array}{ll}
k^2,& \text{for $a=0$}\,,
\cr
1,& \text{for $b=0$}\,.
\end{array}
\right.
}
The relation for masses squared depends on mixing angles as
\eq{
M_1^2+M_2^2+M_3^2=m_D^4\Big[\kappa^4s(m_i^{-2})+4|d|^2\kappa^2(1-\kappa^2)\Big]
+
\left\{\begin{array}{ll}
0,& \text{for $a=0$}\,,
\cr
m_D^4(1-\kappa^4)s^2(m_i^{\prime -1}),& \text{for $b=0$}\,,
\end{array}
\right.
}
where we used the shorthand
\eq{
s(a_i)=a_1+a_2+a_3,
}
and $s^2(a_i)=\big(s(a_i)\big)^2$.
The value of $|d|$ also depends on light masses and mixing angles, 
cf.\,\eqref{params},
\eq{
|d|^2=\ums{2}c^2_{13}\left[
\left(\frac{1}{m_1'}-\frac{1}{m_2'}\right)^2s^2_{12}c^2_{12}
+\left(\frac{c^2_{12}}{m_1'}+\frac{s^2_{12}}{m_2'}
-\frac{1}{m_3'}
\right)^2s^2_{13}
\right]
\,.
}

\section{Parametrization for $\Uz_R$}
\label{ap:UR}

Here we show how to recover $\Uz_R$ parametrized as \eqref{U0} from the knowledge of only two of the first entries of its first row\,\cite{mutau-r:HS}.

The parametrization in \eqref{U0} is
\eq{
\Uz_R=\mtrx{u_1&u_2&u_3\cr
	w_1& w_2 & w_3\cr
	w_1^*& w_2^* & w_3^*}\,,
}
where we can decompose $w_k$ as
\eq{
w_k=|w_k|e^{i\gamma_k}\,.
}

The modulus and relative phases of the second and third rows can be obtained from orthogonality as
\eqali{
|w_i|&=\sqrt{\frac{1-u_i^2}{2}}\,,
\cr
\cos\gamma_{ij}&=-\frac{u_iu_j}{\sqrt{1-u_{i\phantom{j}}^2\!}\sqrt{1-u_j^2}}
\,,
}
where $\gamma_{ji}\equiv\gamma_j-\gamma_i$.
The quadrant ambiguity of $\gamma_{ij}$ can be resolved by either one of the unitary triangles
\eqali{
|u_1w_1|+|u_2w_2|e^{i\gamma_{21}}+|u_3w_3|e^{i\gamma_{31}}&=0\,,
\cr
|w_1|^2+|w_2|^2e^{i2\gamma_{21}}+|w_3|^2e^{i2\gamma_{31}}&=0\,.
}
The individual $\gamma_i$ are most easily calculated in the phase convention where $D$ 
in \eqref{MR:param} is real and recall that $\Uz_R$ diagonalizes $M_R^*$ in our convention.
In this case, the eigenvector equation leads to 
\eq{
\cos\gamma_i=\frac{M_i'-A}{D}\frac{u_i}{\sqrt{2(1-u_i^2)}}\,,
\quad
Du_i\sin\gamma_i=-|w_i|(B\sin2\gamma_i+\im C)\,.
}
Rephasing $D\to e^{i\alpha}D$ modifies $w_i\to e^{i\alpha} w_i$, i.e., $\gamma_i\to 
\gamma_i+\alpha$.

\section{Simplified CP asymmetry}
\label{ap:eps}

The full expression of the simplified CP asymmetry \eqref{epstau:simp} in the $\tau$ flavor is
\eq{
\label{epstau:simp:full}
\eps^{(0)}_\tau=-\frac{\yN{e}^2}{8\pi}\frac{\kappa^2(1-\kappa^2)}{\kappa^2+(1-\kappa^2)|\Uz_{Re0}|^2}
\sum_{j\neq 0}
\im\big[{\Uz_{Re 0}}^*\Uz_{Re j}\Uz_{R\tau 0}{\Uz_{R\tau j}}^*
\big]
\Big[K_{R00}^2K_{Rjj}^2g(x_j)+\frac{1}{1-x_j}
\Big]\,.
}
We describe briefly in the following how to obtain it.
We stress that there is no resonant enhancement if $K_{R00}^2K_{Rjj}^2=-1$, i.e., if $\cN_0$ and $\cN_j$ have opposite CP parity, because the combination $-g(x)+1/(1-x)$ approaches $-1/2+\ln(4)\approx 0.88$ in the limit $x\to 1$.

We use the fact that the Yukawa coupling \eqref{lambda:k} is proportional to an unitary matrix when $\kappa=1$.
Away from that point we can rewrite \eqref{lambda:k} as
\eq{
\lambda=\yN{e} U_R^\dag[\kappa\id +(1-\kappa)e_{11}]\,,
}
where $e_{11}=\diag(1,0,0)$ is the projector into $e$ flavor.
For definiteness we also assume the lightest heavy neutrino is $\cN_1$ as conventionally  adopted.
We also use the decomposition for $U_R$ in \eqref{U=U0.K}.
These properties allows us to rewrite the combinations
\eqali{
\lambda_{j\alpha}\lambda_{1\alpha}^* &=\yN{e}^2(K_R)_{11}(K_R)_{jj}^*
\Big[
\kappa^2(\Uz_{R})_{\alpha 1}(\Uz_{R})_{\alpha j}
+(1-\kappa^2)(\Uz_{R})_{e 1}(\Uz_{R})_{e j}^*
\Big]\,,
\cr
(\lambda\lambda^\dag)_{j1}&=\yN{e}^2
\Big[
\kappa^2\delta_{1j}+(1-\kappa^2)(\Uz_{R})_{e 1}(\Uz_{R})_{e j}^*
(K_R)_{11}(K_R)_{jj}^*
\Big]\,,
\cr
(\lambda\lambda^\dag)_{1j}&=\yN{e}^2
\Big[
\kappa^2\delta_{1j}+(1-\kappa^2)(\Uz_{R})_{e j}(\Uz_{R})_{e 1}^*
(K_R)_{jj}(K_R)_{11}^*
\Big]\,.
}
Certain combinations become real and we are only left, for $\alpha=\tau$, with
\eqali{
\im\big[(\lambda\lambda^\dag)_{j1}\lambda_{j\tau}\lambda^*_{1\tau}\big]
&=
\yN{e}^4\kappa^2(1-\kappa^2)(K_R)^2_{11}(K_R)^2_{jj}
\im\big[(\Uz_{R})_{e 1}(\Uz_{R})_{e j}^*(\Uz_{R})_{\tau 1}(\Uz_{R})_{\tau j}^*
\big]\,,
\cr
\im\big[(\lambda\lambda^\dag)_{1j}\lambda_{j\tau}\lambda^*_{1\tau}\big]
&=
\yN{e}^4\kappa^2(1-\kappa^2)
\im\big[{\Uz_{Re 1}}^*\Uz_{Re j}\Uz_{R\tau 1}{\Uz_{R\tau j}}^*
\big]\,.
}


\end{document}